\documentclass[5p,times]{elsarticle}
\usepackage{hyperref}

\journal{Information Science} 

\usepackage{balance}
\usepackage{listings}
\usepackage{color}

\definecolor{dkgreen}{rgb}{0,0.6,0}
\definecolor{gray}{rgb}{0.5,0.5,0.5}
\definecolor{mauve}{rgb}{0.58,0,0.82}

\makeatletter
\def\lst@makecaption{%
  \def\@captype{table}%
  \@makecaption
}
\makeatother

\graphicspath{{img/}}

\usepackage{url}
\usepackage{multirow}

\usepackage{amssymb}
\usepackage{amsmath}
\newtheorem{definition}{Definition}
\newtheorem{strategy}{Strategy}    
\newtheorem{theorem}{Theorem}   
\usepackage{algorithm}
\usepackage{algorithmic}
\usepackage{array}

\usepackage[caption=false,font=footnotesize,labelfont=sf,textfont=sf]{subfig}

\begin{document}

\begin{frontmatter}

\title{ProUM: Projection-Based Utility Mining on Sequence Data}

\author{Wensheng Gan$ ^{1,5} $,
	Jerry Chun-Wei Lin$ ^{1,2}$*,
	Jiexiong Zhang$ ^{1} $,
	Han-Chieh Chao$ ^{3} $,
	Hamido Fujita$ ^{4} $ and
	Philip S. Yu$ ^{5} $}

\address{$ ^{1} $School of Computer Sciences and Technology, Harbin Institute of Technology (Shenzhen), Shenzhen, Guangdong 518055, China} 
\address{$ ^{2} $Department of Computing, Mathematics and Physics, Western Norway University of Applied Sciences (HVL), Bergen 5050, Norway}
\address{$ ^{3} $Department of Electrical Engineering, National Dong Hwa University, Hualien 97401, Taiwan}
\address{$ ^{4} $Faculty of Software and Information Science,  Iwate Prefectural University, Morioka 020-8550, Japan}
\address{$ ^{5} $Department of Computer Sciences, University of Illinois at Chicago, Chicago, IL 60607, USA}

\address{Email: wsgan001@gmail.com, jerrylin@ieee.org, jiexiong.zhang@foxmail.com, hcc@ndhu.edu.tw, HFujita-799@acm.org, psyu@uic.edu}

\cortext[cor1]{Corresponding author. Email: jerrylin@ieee.org.} 

\begin{abstract}

Utility is an important concept in economics. A variety of applications consider utility in real-life situations, which has lead to the emergence of utility-oriented mining (also called utility mining) in the recent decade. Utility mining has attracted a great amount of attention, but most of the existing studies have been developed to deal with itemset-based data. Time-ordered sequence data is more commonly seen in real-world situations, which is different from itemset-based data. Since they are time-consuming and require large amount of memory usage, current utility mining algorithms still have limitations when dealing with sequence data. In addition, the mining efficiency of utility mining on sequence data still needs to be improved, especially for long sequences or when there is a low minimum utility threshold. In this paper, we propose an efficient \textbf{\underline{Pro}}jection-based \textbf{\underline{U}}tility \textbf{\underline{M}}ining (ProUM) approach  to discover high-utility sequential patterns from sequence data. The utility-array structure is designed to store the necessary information of the sequence-order and utility. ProUM can significantly improve the mining efficiency by utilizing the projection technique in generating utility-array, and it effectively reduces the memory consumption. Furthermore, a new upper bound named sequence extension utility is proposed and several pruning strategies are further applied to improve the efficiency of ProUM. By taking utility theory into account, the derived high-utility sequential patterns have more insightful and interesting information than other kinds of patterns. Experimental results showed that the proposed ProUM algorithm significantly outperformed the state-of-the-art algorithms in terms of execution time, memory usage, and scalability.

\end{abstract}


\end{frontmatter}


\section{Introduction}

In the era of big data, data mining and data analytics \cite{chen1996data} are some of the fundamental technologies for discovering knowledge in data, and they have become more prevalent in our life due to the rapid growth of massive data \cite{4gan2017data}. Up to now, a handful of methods have been proposed for discovering useful and interesting patterns \cite{agrawal1994fast,han2004mining} from different types of data. For example, frequent pattern mining (FPM) \cite{han2004mining} and association rule mining (ARM) \cite{agrawal1994fast} from the transaction data have been extensively studied. One of the well-known applications of FPM and ARM is in market basket analysis. In addition, mining sequence data, which is common for many real-life applications, has also attracted a lot of attention. Sequential pattern mining (SPM) is one of the well-studied research fields for mining sequence data \cite{agrawal1995mining,srikant1996mining,pei2001prefixspan}. Knowing the useful patterns and auxiliary knowledge from sequences/events can benefit a number of applications, such as web access analysis, event prediction, time-aware recommendation, and DNA detection \cite{fournier2017survey}. Up to now, research has been conducted  on mining interesting patterns from transaction or sequential data \cite{fournier2017survey,4gan2017data,han2004mining,pei2001prefixspan}. However, most of them are based on the co-occurrence frequency of patterns.

\textbf{Motivation}. In reality, massive data, for example, a sequence, contains valuable but hidden auxiliary information. However, the measures of support \cite{han2004mining} and confidence \cite{agrawal1995mining,agrawal1994fast} cannot be effectively utilized to discover implicit or potential information.  For instance, implicit factors such as the utility, interest, risk, and profit of objects in the data are not considered in traditional ARM or SPM. The main goal of tasks for data mining and analytics is generally to achieve utility maximization. However, most of the existing algorithms of FPM, ARM, and SPM are unable to discover the valuable targeted patterns that benefit utility maximization. For example, support/frequency-based data mining models might be insufficient for achieving a time-aware recommendation for users based on the users' click-stream or purchase behavior. Specifically, a variety of applications consider utility. Typical examples include the profit of products in supermarkets and retail stores, the satisfaction feedbacks of different restaurants, and the popularity of hot showing movies. Thus, in these circumstances, the frequency framework loses its adaptiveness.

Utility \cite{marshall2005principles} is an important concept in Economics. The emergence of a new mining anc omputing framework, called utility mining, can be realized by taking utility theory \cite{marshall2005principles} from economics into account \cite{2gan2018survey}. Utility mining is in the cross-domain of information technology and economics. In the past decade, utility mining has been extensively studied in areas like high-utility itemset mining (HUIM) \cite{ahmed2009efficient,liu2012mining,liu2005two,tseng2013efficient}, high-utility sequential pattern mining (HUSPM) \cite{alkan2015crom,wang2016efficiently,yin2012uspan}, and high-utility episode mining (HUEM) \cite{wu2013mining}. It has been successfully applied to discover utility-driven knowledge in the cross-domain of information technology and business. The identified patterns, which can bring valuable profits for retailers or managers, are more useful than those frequent-based patterns in business. HUIM addresses the itemset-based data to mine high-utility itemsets (HUIs), and HUSPM deals with the sequence data to discover high-utility sequential patterns (HUSPs). In general, the \textit{utility} can be a user-specified subjective measure, such as satisfaction, profit, risk, interest, and so on. Utility mining \cite{2gan2018survey} has become an important branch of data science, which is aimed at utilizing the auxiliary information from data (e.g., itemsets, events, and sequences). Time-ordered sequence data is more commonly seen in real-world situations, which is different from itemset-based data. To a certain extent, itemset data is a special case of sequence data.

\textbf{Challenges}. Several prior studies have focused on improving the mining efficiency of HUSPM, such as USpan \cite{yin2012uspan}, HuspExt \cite{alkan2015crom}, PHUS \cite{lan2014applying}, and HUS-Span \cite{wang2016efficiently}. Among them, USpan \cite{yin2012uspan}, HuspExt \cite{alkan2015crom}, and HUS-Span \cite{wang2016efficiently} are all based on a lexicographic sequence tree with two concatenation mechanisms and several pruning strategies w.r.t. upper bounds on utility. In general, some challenges remain in addressing the problem of HUSPM, which are described below. 

First, the computing mechanism of the utility of a pattern is different from that of frequency of a pattern. The former is more complicated and the utility of a sequence is not downward closed. This means that the support-based pattern mining techniques and models cannot be directly applied to discover utility-driven patterns, and the pruning search space in HUSPM is more difficult. 

Second, the HUSPM problem is intrinsically more complex than HUIM and FPM. HUSPM may easily face a critical combinatorial explosion of search space without powerful pruning strategies w.r.t. upper bounds on utility because of the inherent time order embedding in sequence data. 

Third, a common way to identify the interesting HUSPs is to recursively generate the projected sub-databases and then scan these whole sub-databases. However, this is very inefficient and costly in memory when the number of  sequences in processed database is large-scale. Therefore, how to efficiently reduce the size of the databases that need to be projected and scanned is a crucial problem to be solved for efficiently discovering HUSPs. 

In summary, speeding up the execution time and reducing memory consumption without losing HUSPs are critical in HUSPM. How to improve the mining efficiency of utility mining on sequence data is still an open problem.

\textbf{Contributions}. In light of the aforementioned challenges, we propose a novel utility mining framework, called  \textbf{\underline{Pro}}jection-based \textbf{\underline{U}}tility \textbf{\underline{M}}ining on sequence data (ProUM). Based on the developed utility-array with the projection mechanism, the utility-driven mining model, ProUM, can not only extract the insightful high-utility patterns but also achieve better efficiency. The effectiveness and efficiency of the proposed ProUM is evaluated by comparing it with the well-known frequency-based SPM model and state-of-the-art utility mining algorithms. The major contributions of this paper can be summarized as follows:

\begin{itemize}

	\item We adopt utility and time-order significance as the key criterion for evaluating utility mining on sequence data. By considering the utility factor and time-order relations among items/objects, we design an efficient method, called \textbf{\underline{Pro}}jection-based \textbf{\underline{U}}tility \textbf{\underline{M}}ining on sequence data (ProUM).
	
	\item  A compact data structure, namely, utility-array, is presented to store the compact information (e.g., utility, position, and time order) of sequences from the processed sequence database. ProUM can quickly discover a set of high-utility sequential patterns based on the developed utility-array with the projection mechanism.  
	
	\item This projection-based approach utilizes several pruning techniques in a depth-first search manner, which consist of the utilization of utility property and the proposed upper bound named sequence extension utility (\textit{SEU}). Therefore, ProUM is able to filter a large number of unpromising patterns at an early stage and return the significant patterns in the mining process. 
	
	\item  Experiments on both real and synthetic datasets show that the proposed utility-array representation achieves a lossless compression capability of quantitative sequence data. Moreover, ProUM significantly outperforms the state-of-the-art algorithms, such as USpan and HUS-Span. 

\end{itemize}

The remainder of this paper is organized as follows: Some related works of support-based sequence mining and utility-based mining on itemset-based data and sequence data are briefly reviewed in Section \ref{sec:relatedwork}. Some basic preliminaries and the problem statement of HUSPM are given in Section \ref{sec:preliminaries}. Details of the proposed data structure, pruning strategies with upper bound, and the main procedure of the ProUM algorithm are described in Section \ref{sec:algorithm}. The experimental evaluation of the proposed ProUM algorithm is provided in Section \ref{sec:experiments}.  Finally, conclusions are drawn in Section \ref{sec:conclusion}.

\section{Literature Review}
\label{sec:relatedwork}

Much research has been conducted on frequency-based sequential pattern mining and utility mining of itemset-based data, but fewer works have integrated utility theory for mining high-utility patterns from event/sequence data. In this section, we separately present the prior works on SPM, HUIM, and HUSPM. 

\subsection{Frequency-Based Mining on Sequences}

Pattern (i.e., itemset, rule, and sequence) mining \cite{han2004mining,pei2001prefixspan} is a kind of well-studied data mining and analytics model. The applications of pattern mining models are very extensive, and details can be referred to in the survey literature \cite{fournier2017survey,4gan2017data,2gan2018survey,3gan2018survey}. A great effort has been put forth by the data mining community to discover frequent patterns from itemset-based data, such as Apriori \cite{agrawal1994fast} and FP-growth \cite{han2004mining} methods. Different from itemset-based data, timely ordered sequence data is more commonly seen in the real-world, in areas such as traffic data, web access, customer shopping data, travel routes, stock market trends, DNA chains, and so on \cite{agrawal1995mining,fournier2017survey,pei2001prefixspan,srikant1996mining}. The problem of sequential pattern mining (SPM) from sequence data was first presented by Agrawal and Srikant \cite{agrawal1995mining}. Frequent pattern mining (FPM) from itemset-based data is closely related to SPM \cite{agrawal1995mining,fournier2017survey,pei2001prefixspan,srikant1996mining}. A number of algorithms have been proposed to discover the complete frequent sequential patterns from sequential databases, including SPADE \cite{zaki2001spade}, SPAM \cite{ayres2002sequential}, and PrefixSpan \cite{pei2001prefixspan}. These algorithms use many strategies to make the mining of sequential patterns more efficient and practical. Unfortunately, before discovering the final result sets, SPM may produce a huge amount of candidates, partially due to the combinatorial nature of the mining task and the timely ordered information embedding in the sequences. Analysis is difficult when dealing with long sequences because most of the SPM algorithms may generate an exponential number of sequences, especially when using a lower minimum support threshold. Among them, the well-known PrefixSpan \cite{pei2001prefixspan} algorithm follows a pattern growth mechanism that uses a series of projected databases for achieving better mining performance in terms of execution time and memory cost. It recursively extracts the  prefix sub-sequences, and then projects the postfix sub-sequences into the sub-databases \cite{pei2001prefixspan}. Comprehensive overviews of sequential pattern mining have been given by Fournier-Viger et al. \cite{fournier2017survey} as well as Gan at al. \cite{3gan2018survey}.

\subsection{Utility-Driven Mining on Transaction Data}
The key measure for discovering patterns in the aforementioned FPM and SPM is the frequency (aka relevant co-occurrence) \cite{han2004mining}. In general, the statistical frequency is an objective measure while some subjective measures and useful factors (e.g., utility, business profit, risk, and preference) are ignored. Therefore, the support/frequency-based data mining approaches cannot return the real useful knowledge, which decreases the effectiveness of mining task. Up to now, a variety of applications have considered utility. A new mining and computing framework named utility mining \cite{2gan2018survey} has been proposed by taking the utility theory from economics into account. Utility mining has been developed to successfully applied to discover utility-driven knowledge in many real-world applications. The early works of utility mining were related to high-utility itemset mining (abbreviated as HUIM) \cite{ahmed2009efficient,liu2012mining,liu2005two,tseng2013efficient}, which addresses itemset-based data. In general, the \textit{utility} can be any user-specified subjective measure, such as satisfaction, profit, risk, interest, and so on.

Many previous studies of utility mining have focused on developing efficient algorithms that can achieve better mining performance, such as Apriori-like approaches (e.g., Two-Phase \cite{liu2005two}), tree-based approaches (e.g., IHUP \cite{ahmed2009efficient}, UP-growth \cite{tseng2010up}, UP-growth+ \cite{tseng2013efficient}), list-based approaches (e.g., HUI-Miner \cite{liu2012mining},  FHM \cite{fournier2014fhm}), and other hybrid algorithms (e.g., EFIM \cite{zida2017efim}). In addition to efficiency, the effectiveness of the data mining and analytic models is also very important. Therefore, a number of studies have been developed to improve the effectiveness for mining utility-oriented patterns, and the current state-of-the-art approaches have been provided in \cite{2gan2018survey}. For instance, Lin et al. studied the problem of dynamic high-utility itemset mining on different types of dynamic data with record insertion \cite{2gan2018survey,lin2015fast}, record deletion \cite{lin2016fast}, and record modification \cite{lin2015fast}. In some applications, the collected data is not precise and contains uncertainty. There have been some interesting works that deal with uncertain data for mining high-utility patterns \cite{lin2016efficient}. At the same time, other interesting issues of utility mining also have been studied, such as utility mining with discount strategies \cite{lin2016fast} or negative values \cite{lin2016fhn}, discovering top-$k$ high-utility patterns \cite{tseng2016efficient}, correlated utility mining \cite{gan2018coupm,gan2017extracting}, and HUIM in big data \cite{lin2015mining}. Recently, a new utility measure, called utility occupancy \cite{gan2017exploiting}, has been proposed to solve the drawbacks of the existing utility mining models.

Different from the above-mentioned approaches, there are several genetic algorithms (e.g., HUIM-BPSO \cite{lin2017binary} and ACO-based HUIM-ACS \cite{wu2017aco}) methods that have been applied to deal with the utility mining problem. However, evolutionary computation techniques for HUIM do not provide any benefit for improving the mining efficiency. Besides, the interesting topic called privacy preserving utility mining \cite{gan2018privacy} also has been extensively studied. A detailed survey of current development of HUIM was reported by Gan et al. \cite{2gan2018survey}.

\subsection{Utility-Driven Mining on Sequences}

In addition to itemset-based data, sequence data also has been addressed in utility mining, which is called high-utility sequential pattern mining (HUSPM) \cite{alkan2015crom,wang2016efficiently,yin2012uspan}. In FPM and SPM, the Apriori property \cite{agrawal1995mining,agrawal1994fast,srikant1996mining} is widely adopted as the downward closure property to prune the search space. However, the Apriori property does not hold in HUSPM, and this makes the analysis of HUSPM difficult. Due to the absence of the downward closure property in sequence utility, the sequence-weighted utilization (\textit{SWU}) \cite{ahmed2010mining,yin2012uspan} is utilized in HUSPM to prune the search space. It has been proved that the \textit{SWU} value is an upper bound of the utilities of a sequence and all its super-sequences. For extracting high-utility sequential patterns, Ahmed et al. \cite{ahmed2010novel} first proposed two algorithms, UtilityLevel and UtilitySpan. UtilityLevel is an Apriori-like algorithm, and UtilitySpan is based on PrefixSpan \cite{pei2001prefixspan}. UL and US discover HUSPs in two phases by using \textit{SWU} to prune the search space. They first find the candidate sequences with a high \textit{SWU} value, then they compute the actual utilities of each sequence in candidates, and finally all the HUSPs can be identified. Next, the UMSP algorithm \cite{shie2011mining} was developed to discover high-utility mobile sequences, and the UWAS-tree and IUWAS-tree \cite{ahmed2010mining} were designed to find high-utility web log sequences.

Unfortunately, all these algorithms only consider single-item sequences (i.e., itemset data that was addressed in HUIM) but not the element-based sequences. Yin et al. \cite{yin2012uspan} presented a generic definition of the HUSP mining framework and proposed a new mining algorithm named USpan. In the USpan model, the quantitative sequences, along with their utility and time order information, are represented as the utility-matrix structure. Then, two upper bounds (\textit{SWU} and sequence-projected utilization (\textit{SPU}) \cite{yin2012uspan}) are applied to prune the search space, which is represented as a lexicographic tree. To prune the lexicographic tree for a better efficiency, \textit{SWU} is utilized in a depth pruning strategy and \textit{SPU} is utilized in a breath pruning strategy. However, the high-utility sequential patterns that are mined by USpan are not complete. Alkan et al. proposed HuspExt \cite{alkan2015crom} with a Cumulate Rest of Match (CRoM) based pruning technique to improve the mining performance. HuspExt also utilizes an upper bound on the utilities of the candidate sequences to prune the search space. Lan et al.  \cite{lan2014applying} further introduced a projection-based PHUS algorithm to mine HUSPs using a sequence-utility upper-bound (\textit{SUUB}) model. The maximum utility measure is introduced in the \textit{SUUB} model to obtain a tighter upper bound on the utility of a sequence.

The above algorithms (e.g., USpan \cite{yin2012uspan} and HuspExt \cite{alkan2015crom}) adopt \textit{SWU} to prune the search space, however, they usually suffer from the problem of an exponential number of candidate sequences, especially when the user-defined minimum utility threshold is small. The HUS-Span algorithm \cite{wang2016efficiently}, which was proposed recently, utilizes a new upper bound, called the prefix extension utility (\textit{PEU}). Although the results discovered by HUS-Span are complete, HUS-Span is not efficient enough. The generate-and-test approach creates an overflow of candidate sequences. Recently, some interesting issues of HUSPM have been extensively studied that can improve the effectiveness of mining high-utility sequential patterns. For example, the problems of mining top-$k$ high-utility sequential patterns \cite{yin2013efficiently,wang2016efficiently}, discovering periodic HUSPs  \cite{dinh2018efficient},  mining HUSPs with multiple minimum utility thresholds \cite{lin2017high}, and incrementally mining HUSPs on a dynamic database \cite{wang2018incremental} have been addressed. It should be noted that several genetic algorithms have been developed for HUIM (e.g., HUIM-BPSO \cite{lin2017binary} and HUIM-ACS \cite{wu2017aco}), but they have not been proposed to deal with HUSPM yet. More current development of HUSPM can be referred to in literature reviews \cite{gan2018privacy,2gan2018survey,truong2019survey}.

\section{Preliminaries and Problem Formulation}
\label{sec:preliminaries}

This section introduces some basic concepts and principles of sequence mining and utility-oriented sequence mining. Some definitions from prior works are adopted to present the problem clearly. More details about the background of sequence data and sequence mining can be found in \cite{fournier2017survey,pei2001prefixspan,zaki2001spade}.

\subsection{Sequence Data}

Let $I$ = \{$i_1, i_2, i_3, \cdots, i_n$\} be a set of distinct items. An \textit{itemset} $X$ is a subset of items, that is, $X \subseteq I $. A sequence $s$ is an ordered list of \textit{itemsets} (also called \textit{elements} or \textit{events}). Note that the items within each element can be unordered, without loss of generality, and it is assumed that they are sorted alphabetically. Additionally, ``$\prec$" is used to represent that one item occurs before another item. In SPM, an item occurs once at most in an element of a sequence. An item can occur multiple times (also called occurred quantity) in an element while in HUSPM. A group of the sequences stored with their identifiers (\textit{sid}) is called a sequence database, denoted as $D$. Thus, a sequence database $D$ = \{$s_1, s_2, \cdots, s_m$\} is a set of sequences/tuples (\textit{sid}, $e_1, \cdots, e_j$), where \textit{sid} is a sequence id and $e_j$ is an element that contains a set of items belonging to $I$.

The total number of items in a sequence is called its length. A sequence with length $k$ is called an $k$-sequence. The size of a sequence is the number of itemsets/elements within this sequence. For example, a sequence $s$ = $<$[\textit{ac}], [\textit{abc}], [\textit{abd}], [\textit{ef}]$>$ consists of six distinct items and four elements. Thus, $s$ is called 10-sequence since the length of $s$ is ten, and its size is four. A \textit{k}-itemset, also called  \textit{k}-\textit{q}-itemset, is an itemset that contains exactly \textit{k} items. A \textit{k}-sequence (\textit{k}-\textit{q}-sequence) is a sequence having \textit{k} items.

\begin{definition} (sub-sequence and sup-sequence)
	\rm Given two sequences, $\alpha$ = $<$$a_1, \cdots, a_m$$>$ and $\beta $ = $<b_1, \cdots, b_n>$, $\alpha$ is called a sub-sequence of $\beta $ iff\footnote{In this paper, the term ``iff" means ``if and only if".} each $a_j$ (1 $ \leq j \leq m$) can be mapped by $b_{i_j}$ $(a_j \subseteq b_{i_j}) $ and preserves the order as 1 $ \leq i_1 < i_2 < \cdots < i_m \leq n $. In other words, if $\alpha$ or $\beta$ contains $\alpha$, then $\beta$ can be called a super-sequence. 
\end{definition}

For example, in considering three sequences, $\alpha$ = $<$[\textit{ac}], [\textit{c}], [\textit{abd}]$>$, $\beta$ = $<$[\textit{ac}], [\textit{d}]$>$, and $\gamma$ = $<$[\textit{cd}]$>$, $\alpha$ is said to be a super-sequence of $\beta$ while $\gamma$ is not a sub-sequence of $\alpha$. The reason for this is that the sequence $\gamma$ = $<$[\textit{cd}]$>$ cannot be mapped to any sequence in $\alpha$.

\begin{definition}(sequence containing)
	\rm  Given two sequences, $ t_1$ and $t_2$, then $t_2$ uniquely contains $t_1$ iff there is only one $ t_1 \subseteq t_2  $  such that $t_1$ = $t_2$, denoted as $ t_1 \sqsubseteq t_2$. Similarly, for a sequence $t$ and a sequence $s$, $s$ uniquely contains $t$, denoted as $ t \sqsubseteq s$, iff there is only one $s'$ and $ s' \subseteq s $ such that $ s' \sim t$.
\end{definition}

A quantitative sequential database (shown in Table \ref{table:db}), is used as a running example in this paper. Table \ref{table:db} has five sequences/transactions and six items. In addition, each item $i_j$ in \textit{D} is associated with a unit utility (also called \textit{external utility}), which is denoted as $ pr(i_{j}) $. The unit utility (e.g., price and profit) for each item is provided in Table \ref{table:profit}, which can be called the \textit{profit-table}. In general, the profit-table is based on the prior knowledge of similar users or contents. In the running example, the unit utility of an item (\textit{e}) is \$6.

\begin{table}[!htbp]
	\caption{A quantitative sequence database}
	\label{table:db}
	\centering
	\begin{tabular}{|c|c|}
		\hline
		\textbf{SID} & \textbf{Q-sequence}	 \\ \hline \hline
		$ S_{1} $	& $ < $[(\textit{a}:2) (\textit{c}:1)], [(\textit{c}:2)], [(\textit{b}:10) (\textit{f}:3)], [(\textit{a}:2) (\textit{e}:1)]   $ >$				 \\ \hline
		$ S_{2} $	& $ < $[(\textit{f}:2)], [(\textit{a}:5) (\textit{d}:2)], [(\textit{c}:2)], [(\textit{b}:4)], [(\textit{a}:4) (\textit{d}:1)]$ > $ \\ \hline
		$ S_{3} $	& $ < $[(\textit{a}:4)], [(\textit{b}:4)], [(\textit{f}:5)], [(\textit{a}:1) (\textit{b}:2) (\textit{e}:1)]$ > $			 \\ \hline
		$ S_{4} $	& $ < $[(\textit{a}:3) (\textit{b}:4) (\textit{d}:5)], [(\textit{c}:2) (\textit{e}:1)]$ > $	 \\ \hline 
		$ S_{5} $	& $ < $[(\textit{b}:1) (\textit{e}:1)], [(\textit{c}:1)], [(\textit{f}:2)], [(\textit{d}:2)], [(\textit{a}:4) (\textit{e}:2)]$ > $	 \\ \hline 
	\end{tabular}
\end{table}

\begin{table}[!htbp]
	\caption{A profit-table}
	\label{table:profit}
	\centering
	\begin{tabular}{|c|c|c|c|c|c|c|}
		\hline
		\textbf{Item}	    & \textit{a}	& \textit{b}	& \textit{c}	& \textit{d}	& \textit{e}	& \textit{f} \\ \hline \hline
		\textbf{Profit}	& \$3 & \$2	& \$10 & \$4 & \$6 & \$1 \\ \hline
	\end{tabular}
\end{table}

\begin{definition} (quantitative sequence)
	\rm  For the addressed HUSPM problem, the processed database is the quantitative sequence database (\textit{q}-database) that each item $i_j \in I$ (1 $\leq j \leq n$) in an element/itemset $v$ is associated with a \textit{quantity} (also called \textit{internal utility}), denoted as $q(i_j,v)$. For convenience, ``$q$-" is used to refer to the object associated with quantity throughout this paper. Thus, the term ``\textit{q}-sequence" means a sequence with quantities, and ``sequence" means a sequence without quantities.  Similarly, the ``\textit{q}-itemset" means an itemset having quantities while an ``itemset" does not have quantities.  
	
\end{definition}

\subsection{Utility Mining on Sequence Data}

Utility mining incorporates the utility theory and mining techniques to deal with complex data, such as quantitative sequence data. Some definitions in the utility framework on sequence data are briefly introduced below. 

\begin{definition}(utility of \textit{q}-item)
	\rm  Let $ q(i_{j}, v) $ be the quantity of ($ i_{j} $) in  a \textit{q}-itemset $ v $, and $ pr(i_{j}) $ be the unit profit of ($ i_{j} $). The utility of a \textit{q}-item ($ i_{j} $) in a \textit{q}-itemset \textit{v} is denoted as $ u(i_{j}, v) $ and defined as:
	\begin{equation}
	u(i_{j}, v) = q(i_{j}, v)\times pr(i_{j}).
	\end{equation}
\end{definition}

\begin{definition}(utility of \textit{q}-itemset)
	\rm  The utility of a \textit{q}-itemset $ v $ is denoted as $ u(v) $ and defined as:
	\begin{equation}
	u(v) = \sum_{i_{j}\in v}u(i_{j}, v).
	\end{equation}
\end{definition}

\begin{definition}(utility of \textit{q}-sequence)	
	\rm  The utility of a \textit{q}-sequence $ s $ = $<$$v_{1}, v_{2}, \cdots, v_{d}$$>$ is defined as:
	\begin{equation}
	u(s) = \sum_{v\in s}u(v).
	\end{equation}
\end{definition}

\begin{definition}(utility of \textit{q}-database)
	\rm  The utility of a quantitative sequential database \textit{D} is the sum of the utility of each of its q-sequences:
	\begin{equation}
	u(D) = \sum_{s\in D}u(s).
	\end{equation}			
\end{definition}

For instance, consider the running example in Table \ref{table:db}. The utility of the item (\textit{a}) in the first \textit{q}-itemset in $ S_{1} $ is calculated as: \textit{u}($a$, [(\textit{a}:2) (\textit{c}:1)])  = \textit{q}($a$, [(\textit{a}:2) (\textit{c}:1)]) $\times $ $pr(a)$ = 2 $\times$ \$3 = \$6. In addition, the utility of the first \textit{q}-itemset $<$[(\textit{a}:2) (\textit{c}:1)]$>$ is  \textit{u}([(\textit{a}:2) (\textit{c}:1)]) = \textit{u}(\textit{a}, [(\textit{a}:2) (\textit{c}:1)]) + \textit{u}(\textit{c}, [(\textit{a}:2) (\textit{c}:1)]) = 2 $\times$ \$3 + 1 $\times$ \$10 = \$16.
We have that $ u(S_{1}) $ = \textit{u}([(\textit{a}:2) (\textit{c}:1)]) +
\textit{u}([(\textit{c}:2)]) + \textit{u}([(\textit{b}:10) (\textit{f}:3)]) + \textit{u}([(\textit{a}:2) (\textit{e}:1)]) = \$16 + \$20 + \$23 + \$12 = \$71. Therefore, the overall utility in Table \ref{table:db} is  $ u(D) $ = $ u(S_{1}) $ + $ u(S_{2}) $ + $ u(S_{3}) $ + $ u(S_{4}) $ + $ u(S_{5}) $ = \$71 + \$69 + \$38 + \$63 + \$52 = \$293.

\begin{definition}(sequence matching)
	\rm  Given a \textit{q}-sequence \textit{s} = $<$$v_{1},$ $v_{2},$ $\cdots, v_{d}$$>$ and a sequence $ t $ = $<$$w_{1}, w_{2}, \cdots, w_{d'}$$>$, if $ d $ = $ d' $ and the items in $ v_{k} $ are the same as the items in $ w_{k} $ for $ 1\leq k\leq d $, then $ t $ matches $ s $, which is denoted as $ t\sim s $.	
\end{definition}

For instance, in Table \ref{table:db}, the sequences $<$[\textit{ac}]$>$, $<$[\textit{ac}], [\textit{b}]$>$, $<$[\textit{a}], [\textit{b}], [\textit{e}]$>$ all match $ S_{1} $. A sequence in a $q$-sequence database may have more than one match in a $q$-sequence. For instance, $<$[\textit{a}], [\textit{b}]$>$ has two matches in $ S_{3} $, such as $<$[\textit{a}:4], [\textit{b}:4]$>$ and  $<$[\textit{a}:4], [\textit{b}:2]$>$. The measure of the utility of sequences for HUSPM is more challenging than that for SPM and HUIM due to the multiple matching cases.

\begin{definition}($q$-itemset containment)
	\rm  Given two itemsets $ w $ and $ w' $, the itemset $ w $ is contained in $ w' $ (denoted as $ w\subseteq w' $) if $ w $ is a subset of $ w' $ or $ w $ is the same as $ w' $. 
	Given two \textit{q}-itemsets $ v $ and $ v' $, $ v $ is said to be contained in $ v' $ if for any item in $ v $ there exists the same item having the same quantity in $ v' $. This is denoted as $ v\subseteq v' $. 
\end{definition}

\begin{definition}($q$-sequence containment)
	\rm  Given two sequences \textit{t} = $<$$w_{1}, w_{2}, \cdots, w_{d}$$>$ and $ t' $ = $<$$w'_{1}, w'_{2}, \cdots, w'_{d'}$$>$, the sequence $ t $ is contained in $ t' $ (denoted as $ t\subseteq t $') if there exists an integer sequence $ 1\leq k_{1}\leq k_{2}\leq \cdots \leq d' $ such that $ w_{j}\subseteq w'_{k_{j}} $ for $ 1\leq j\leq d $. 
	In addition, consider two \textit{q}-sequences $ s $ = $<$$v_{1}, v_{2}, \cdots, v_{d}$$>$ and $ s' $ = $<$$v'_{1}, v'_{2}, \cdots, v'_{d'}$$>$. We say $ s $ is contained in $ s' $ (denoted  as $ s\subseteq s' $) if there exists an integer sequence $ 1\leq k_{1}\leq k_{2} \leq \cdots \leq d' $ such that $ v_{j}\subseteq v'_{k_{j}} $ for $ 1\leq j \leq d $.
	In the following, $ t \subseteq s $ is used to indicate that $ t \sim s_{k} \wedge s_{k} \subseteq s $ for convenience.
\end{definition}

For example,  the itemset [\textit{ab}] is contained in the itemset [\textit{abe}], while [\textit{abe}] does not contain the item [\textit{f}]. The \textit{q}-itemset [(\textit{a}:1) (\textit{b}:2)] is contained in [(\textit{a}:1) (\textit{b}:2) (\textit{e}:1)], but it is not contained in [(\textit{a}:3) (\textit{b}:4) (\textit{d}:5)]. In Table \ref{table:db}, $S_3$ contains [(\textit{a}:1) (\textit{b}:2)], but $S_4$ does not contain it. Consequently, the sequences $<$[(\textit{a}:4)], [(\textit{b}:4)]$>$ and $<$[(\textit{a}:4)], [(\textit{b}:2)]$>$ are contained in $ S_{3} $, but $<$[(\textit{a}:3)], [\textit{b}:2]$>$ is not contained in $ S_{3} $.

It should be noted that the definition of utility of sequence originally proposed in \cite{ahmed2010mining,ahmed2010novel} is too specific. Therefore, the later studies \cite{wang2016efficiently,yin2012uspan} have adopted ``the maximum utility of all occurrences of a sequence $t$ in a $q$-sequence $s$" as the real utility of $t$ in $s$. The proposed model in this paper follows this definition of utility. 

\begin{definition}(maximal utility of $ t $ in \textit{s})
	\rm  Consider a sequence $t$ and a $q$-sequence $s$. The utility of $ t $ in \textit{s}, denoted as $u(t, s)$, may have different utility values. The maximum utility is chosen among these utility values as the utility of $t$ in $s$, as defined below:
	\begin{equation}
		u(t, s) = max\{u(s_{k})|t \sim s_{k} \wedge s_{k} \subseteq s\}.
	\end{equation}
\end{definition}

\begin{definition}(utility of a sequence in $D$)
	\rm  Let $ u(t) $ denote the overall utility of a sequence $ t $ in a quantitative sequential database $ D $. It is defined as:
	\begin{equation}
		u(t) = \sum_{t \subseteq s \wedge s\in D} u(t,s).
	\end{equation}
\end{definition}

For instance, consider two sequences $<$$[a]$, $[b]$$>$ and $<$$[f]$, $[ad]$$>$ in Table \ref{table:db}. $<$$[a]$, $[b]$$>$ has two utility values in $S_3$, and thus $ u$($<$$[a]$, $[b]$$>$, $S_{3}) $ = $ max$\{\textit{u}($<$[\textit{a}:4], [\textit{b}:4]$>$), \textit{u}($<$[\textit{a}:4], [\textit{b}:2]$>$)\} = $ max $\{\$20, \$16\} = \$20. $<$$[f],[ad]$$>$ also has two utility values in $S_2$, such that  $ u$($<$$[f],[ad]$$>$, $S_{2}) $ = $max$\{\textit{u}($<$[\textit{f}:2], [(\textit{a}:5) (\textit{d}:2)]$>$), \textit{u}($<$[\textit{f}:2], [(\textit{a}:4) (\textit{d}:1)]$>$)\} = $ max $\{\$25, \$18\} = \$25. Intuitively, the calculation of the overall utility of a sequence in the sequence database is quite a bit more complicated than that of HUIM and SPM.

Therefore, the overall utility of $<$$[a],[b]$$>$ in Table \ref{table:db} can be obtained as \textit{u}($<$[\textit{a}], [\textit{b}]$>$) = \textit{u}($<$[\textit{a}], [\textit{b}]$>$, $S_{1}$) + \textit{u}($<$[\textit{a}], [\textit{b}]$>$, $S_{2}$) + \textit{u}($<$[\textit{a}], [\textit{b}]$>$, $S_{3}$) = \$26 + \$23 + \$20 = \$69. Notice that $<$$[a],[b]$$>$ is different for $<$$[b],[a]$$>$ because HUSPM considers the time orders embedding in sequences. Thus, $<$$[b],[a]$$>$ is contained in $S_5$ while $<$$[a],[b]$$>$ is not contained in $S_5$.

\subsection{Problem Definition}

\begin{definition}(high-utility sequential pattern, HUSP)
	\rm In a quantitative sequential database \textit{D}, a sequence $ t $ is said to be a high-utility sequential pattern (denoted as \textit{HUSP}) if its overall utility in $D$ satisfies:
	\begin{equation}
		HUSP \gets \{t|u(t)\geq \delta \times u(D)\}.
	\end{equation}
	where $\delta$ is the minimum utility threshold $\delta$ (usually given as a percentage).
\end{definition}

In the running example, it is assumed that $ \delta$ is set as 25\%, and then  $ \delta $ $\times u(D) $ =  25\% $\times $ \$293 =  \$73.25. Thus, since \textit{u}($ < $[\textit{a}], [\textit{b}]$ > $) = \$69 $ < $ \$73.25, $ < $[\textit{a}], [\textit{b}]$ > $ is not a HUSP. Based on the above-stated concepts, the formal definition of the utility mining on sequence data (also called high-utility sequential pattern mining) problem can be defined below.

\textbf{Problem Statement:}  Given a quantitative sequential database $D$ (with a profit-table) and a user-defined minimum utility threshold $\delta$, the utility-driven mining problem of high-utility sequential pattern mining (HUSPM) consists of enumerating all HUSPs whose overall utility values in this database are no less than the prespecified minimum utility account, such as $\delta \times u(D) $. 

Therefore, the goal of HUSPM is to search for the set of sequences that achieves the highest utility score and their utility values are not less than the  minimum utility value.


\section{Proposed Utility Mining Algorithm: ProUM}
\label{sec:algorithm}

\begin{figure}[!htbp]
	\setlength{\abovecaptionskip}{0pt}
	\setlength{\belowcaptionskip}{0pt}	
	\centering
	\includegraphics[scale=0.7]{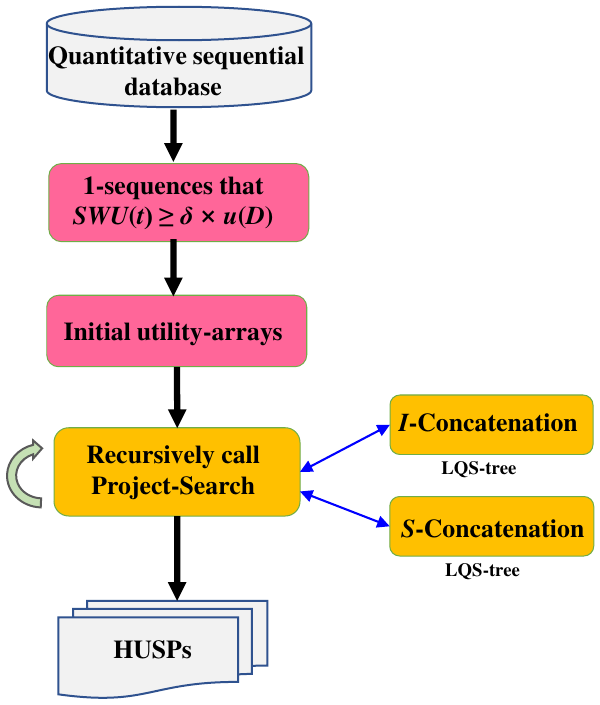}
	\caption{Framework of the ProUM algorithm}
	\label{fig:framework}
\end{figure}

This section describes the proposed projection-based ProUM algorithm for discovering high-utility sequence-based patterns by recursively projecting the utility-array based on the prefix sequences. ProUM utilizes the utility-array data structure to avoid multiple scans of the original database and projecting of the sub-databases. Only the compact utility-array is needed to be projected and scanned in each mining process. The framework of the proposed ProUM algorithm is presented in Figure \ref{fig:framework}. First, details of the search space, the utility-array structure, and the projection mechanism are presented below.

\subsection{Lexicographic Sequence Tree}

According previous studies (e.g., SPAM \cite{ayres2002sequential}, PrefixSpan \cite{pei2001prefixspan}, USpan \cite{yin2012uspan}), and the complete search space of SPM and HUSPM can be represented abstractly as the lexicographic sequence tree \cite{ayres2002sequential}. For the addressed problem for mining high-utility sequential patterns, a lexicographic $q$-sequence tree (LQS-tree) that was used in USpan \cite{yin2012uspan} is adopted to present the search space of ProUM.  

\begin{definition}(lexicographic $q$-sequence tree)
	\rm  A lexicographic $q$-sequence tree (LQS-tree) is a variant of the lexicographic tree structure \cite{ayres2002sequential} satisfying the following conditions:
	\begin{itemize}
		\item Each node in LQS-tree is associated with a $q$-sequence along with the utility of the sequence while the root is empty and labeled with ``$<$$root$$>$".
		\item All the nodes except the root in LQS-tree are the prefix-based extension of its parent node; in other words, any node's child is an extension sequence/node of the node itself. 
		\item  All the children of any node in LQS-tree are listed in an increasing order, for example, lexicographic order.
	\end{itemize}
\end{definition}

It should be noted that the LQS-tree is just a conceptual structure, and the real visited/searched space may be different depending on different cases.  Additionally, if $\delta $ = 0, then the complete set of the found high-utility sequential patterns from the sequence data is equal to a complete LQS-tree, which covers the complete search space. It is important to notice that the utility property in LQS-tree is different from the frequent/support property in the previous lexicographic tree \cite{ayres2002sequential,pei2001prefixspan}.

\begin{definition}(\textit{I}-\textit{Concatenation} and \textit{S}-\textit{Concatenation})
	\rm There are two operations, called \textit{I}-\textit{Concatenation} and \textit{S}-\textit{Concatenation}, to generate new sequences based on prefix node in LQS-tree. 
	 \begin{itemize}
	 	\item Given a sequence $ t $ and an item $ i_{j} $, the \textit{I-Concatenation} of $ t $ with $ i_{j} $ consists of appending $ i_{j} $ to the last itemset of $ t $, denoted as $<$$t \oplus i_{j}$$>$$_{I-Concatenation}$. 	 
	    \item  An \textit{S-Concatenation} of $ t $ with an item $ i_{j} $ consists of adding $ i_{j} $ to a new itemset appended after the last itemset of $ t $, denoted as $<$$t \oplus i_{j}$$>$$_{S-Concatenation}$.
	\end{itemize}
\end{definition}

For example, given a sequence $ t $ = $<$[$a$], [$c$]$>$ and a new item $(d)$, $<$$t \oplus c$$>$$_{I-Concatenation}$ = $<$[\textit{a}], [\textit{cd}]$>$ and $<$$t \oplus c$$>$$_{S-Concatenation}$ = $<$[\textit{a}], [\textit{c}], [\textit{d}]$>$, then based on the definition of sequence length and sequence size, the \textit{I}-\textit{Concatenation} does not increase the length of sequence $t$ while the \textit{S-Concatenation} increases the length of sequence $t$ since the number of itemsets in $ t $ increases by one. All the candidate sequences in the search space w.r.t. LQS-tree can be enumerated for the purpose of mining HUSPs based on the two prefix-based operations.

Without loss of generality, it is assumed that the concatenated nodes in LQS-tree are listed in alphabetical order. In the running example in Table \ref{table:db}, the sequence in $S_1$ is listed as $ < $[(\textit{a}) (\textit{c})], [(\textit{c})], [(\textit{b}) (\textit{f})], [(\textit{a}) (\textit{e})]   $ >$ instead of $ < $[(\textit{c}) (\textit{a})], [(\textit{c})], [(\textit{f}) (\textit{b})], [(\textit{e}) (\textit{a})]. The expression of a sequence is unique using such a convention.

It should be noted that each LQS-tree node represents a candidate of the search space of HUSPs. A part of the LQS-tree can be referred to in \cite{lin2017high,yin2012uspan}. All the child nodes of a parent node are assumed to be ordered lexicographically with $I$-extension\footnote{Notice that the terms \textit{I}-\textit{Concatenation} and $I$-extension are used interchangeably in this paper} sequences before the $S$-extension\footnote{Notice that the terms \textit{S}-\textit{Concatenation} and $S$-extension are used interchangeably in this paper} sequences. The most straightforward way for spanning the tree structure is to traverse and determine the lexicographic tree node one-by-one using either the Depth-First-Search (DFS) strategy or the Breadth-First-Search (BFS) strategy. Three questions remain to be answered: (1) how to compact the necessary information (e.g., utility, order of sequences, and position in each sequence) from the sequence database into the LQS-tree; (2) how to effectively traverse the nodes in LQS-tree and then quickly calculate their utility values; and (3) how to effectively prune the search space without spanning the compete LQS-tree. To address these problems, we propose a new data structure, utility-array, and several pruning strategies, which are respectively described as the following subsections.

\subsection{Utility-Array and Projection Mechanism}

Real sequences may often be very long in some application scenarios; for example, web search sequences, DNA, network intrusion access/log, and numerous other sequence data are all complex and long. This may easy lead to a huge search space for HUSPM. By adopting the concept of remaining utility \cite{yin2012uspan,wang2016efficiently}, we propose a new compact data structure, called utility-array, for storing the necessary information from the sequence data, including the utility information of item/sequence as well as their time/sequence order.

\begin{definition}(remaining utility \cite{yin2012uspan,wang2016efficiently})
	\rm  Given a sequence $t$ and a sequence database $D$, the remaining utility of $t$ in a $q$-sequence $s$ is the overall utilities of all items whose positions are after $t$ in $s$, and defined as: $u_{rest}(t,s)$ =  \textit{max}\{$u_{rest}(t, p_k, s)$\}, where $p_k$ is the $k$-position, and $u_{rest}(t, p_k, s)$ = $\sum_{i'\in s \wedge t \prec i'}u(i')$. Thus, the overall remaining utility of $t$ in $D$ is defined as: $u_{rest}(t)$ =  $\sum_{s\in D}u_{rest}(t,s)$. 
\end{definition}

Basically, the remaining utility of a sequence means the sum of the utilities after this sequence. Intuitively, the remaining utility is based on matching position. For example, a sequence $<$$s$$>$ has two remaining utility values in $S_1$: $ u_{rest}($$<$$a$$>$$, p_1, s_1) $ = \$10 + \$20 + (\$20 + \$3) + (\$6 + \$6) = \$65 and $ u_{rest}($$<$$a$$>$$, p_2, s_1) $ = \$6. In addition, the remaining utility of $<$$[a]$, $[b]$$>$ in $s_1$ is $ u_{rest}($$<$$[a]$, $[b]$$>$$, s_1) $ = \$3 + (\$6 + \$6) = \$15 and $ u_{rest}($$<$$[a]$, $[b]$$>$$)$ = $ u_{rest}($$<$$[a]$, $[b]$$>$$, s_1) $ + $ u_{rest}($$<$$[a]$, $[b]$$>$$, s_2) $ + $ u_{rest}($$<$$[a]$, $[b]$$>$$, s_3) $ = \$15 + \$16 + \$36 = \$67. Based on the concept of remaining utility, we introduce a data structure to represent the necessary information (both the utility and sequence order w.r.t. position) of each $q$-sequence.

\begin{definition} (utility-array)
	\rm Suppose that all the items in a $q$-sequence $s$ in a $q$-sequence database have different unique occurred positions are \{$p_1$, $p_2$, $ \cdots $, $p_k$\}, where \{$p_1$ $< p_2$ $< \cdots < p_k$\}, and the total number of positions is equal to the length of $s$. The utility-array of a $q$-sequence $s$ = $\langle e_1, e_2, \cdots, e_n \rangle$ ($e_n$ is the $n$-element in $s$) consists of a set of arrays from left to right in $s$. Each array is related to an item $i_j$ in each position $p_k$ and contains the following fields: \textit{array}$_{p_k}$ = $[$\textit{eid}, \textit{item}, \textit{u}, \textit{ru}, \textit{next\_pos}, \textit{next\_eid}$]$. Details are given below:
	\begin{itemize}
		\item  Field \textit{eid} is the element ID of an element containing $i_j$;
		\item  Field \textit{item} is the  name of item $i_j$;
		\item  Field $u$ is the actual utility of $i_j$ in position $p_k$;
		\item  Field $ru$ is the remaining utility of $i_j$ in position $p_k$;
		\item  Field \textit{next\_pos} is the next position of $i_j$ in $s$;
		\item  Field \textit{next\_eid} is the position of the first item in next element (\textit{eid}+1) after current element (\textit{eid}).
	\end{itemize}
	In addition, the utility-array records the first occurred  position of each distinct item in $s$. In summary, a utility-array of a $q$-sequence $s$ is a set of arrays related to each item in $s$ and contains the position, utility, and sequence order information. 
\end{definition}

In the running example in Table \ref{table:db}, the constructed utility-array can be described in Table \ref{table:utilityarray}, and it can be seen that the first occurring position of each distinct item in $s$ is also recorded in the utility-array, for example, the first occurring positions of $a$ and $f$ are 1 and 5, respectively. Intuitively, the utility-array contains all the necessary information of each sequence $t$, including not only the utility values of each item in each position/element but also the sequence order and position information\footnote{Note that ``-" means the position is empty.}.

\begin{table}[!htbp] 
	\setlength{\abovecaptionskip}{0pt}
	\setlength{\belowcaptionskip}{0pt}
	\caption{The utility-array structure of $ S_{1} $}.
	\label{table:utilityarray}
	\centering
	\begin{tabular}{|c|c|c|c|c|c|c|}
		\hline
		 &  \textit{\textbf{eid}}  & \textit{\textbf{item}}  &  \textit{\textbf{u}}  &  \textit{\textbf{ru}}  &  \textit{\textbf{next\_pos}}  &  \textit{\textbf{next\_eid}} \\ \hline \hline
	\textit{array$_1$} & 1   & $a$  & \$6  &  \$65 &  6  & 3 \\ \hline
	\textit{array$_2$} & 1   & $c$  & \$10  &  \$55 &  3  & 3 \\ \hline
	\textit{array$_3$} & 2   & $c$  & \$20  &  \$35 &  -  & 4 \\ \hline
	\textit{array$_4$} & 3   & $b$  & \$20  &  \$15 &  -  & 6 \\ \hline
	\textit{array$_5$} & 3   & $f$  & \$3  &  \$12 &  -  & 6 \\ \hline
	\textit{array$_6$} & 4   & $a$  & \$6  &  \$6 &  -  & - \\ \hline
	\textit{array$_7$} & 4   & $e$  & \$6  &  \$0 &  -  & - \\ \hline
		
	\end{tabular}
\end{table}

\begin{definition} (position in utility-array)
	\rm Each array has a unique position as an index in the designed utility-array structure. Moreover, the size of the arrays in the utility-array of a $q$-sequence $s$ is equal to the length of $s$. This position represents a match of an item $i_j$ in $s$ and can be used as an index for quickly retrieving the utility-array and calculating the detailed information of $i_j$.
\end{definition}

In the implementation details, an array is used to store the set of information of the compact utility-array. Thus, position $pos_k$ indexes \textit{array}$_{pos_k}$. For example, in Table \ref{table:utilityarray}, the \textit{array}$_1$ is indexed by position 1, and the \textit{array}$_2$ is indexed by position 2. First, the 1-sequences with the low \textit{SWU} values in each sequence are removed when calculating the compact utility-array of original sequence database $D$. Then, the new (revised) transactions are used to construct the initial utility-arrays. Inspired by the database projection idea of PrefixSpan \cite{pei2001prefixspan}, we present the following prefix-projected and span mechanism in utility-array.

\begin{definition}(prefix, suffix, and projection \cite{pei2001prefixspan})
	\rm  Assume all the items in an element of a sequence database \textit{D} are listed alphabetically. Given two sequences,  $\alpha$ = $<$$e_{1}$, $e_{2}$, $\cdots$, $e_{n}$$>$,  and $\beta$ = $<$$e'_{1}$, $e'_{2}$, $\cdots$, $e'_{m}$$>$ ($m \leq n$), $\beta$ is called a \textit{prefix} of  $\alpha$ iff it meets the following conditions: (1) $e'_{i} =e_{i} $ for $i \leq m-1$; (2) $ e'_{m} \subseteq e_m$; and (3) all the items in ($e_{m} - e'_{m}$) are alphabetically after those in $e'_{m}$. Additionally, the remaining part/elements after the prefix $\beta$ in a sequence are called \textit{suffix} with regards to prefix $\beta$. Let $\alpha$ be a sequence in $D$, then the $\alpha$-\textit{projected sub-database}, denoted as $D|_\alpha$, is the collection of suffixes of the sequences (which contains $\alpha$) in \textit{D} with regards to prefix $\alpha$. 
\end{definition}

The ProUM algorithm recursively partitions the processed utility-array based on the projection mechanism \cite{pei2001prefixspan}. Specifically, each subset of the extracted sequential patterns are further divided when necessary. Thus, the projection mechanism \cite{pei2001prefixspan} forms a divide-and-conquer framework. Correspondingly, ProUM recursively constructs the corresponding projected utility-arrays but not the projected sub-databases.

\begin{definition} (projected utility-array)
	\rm Let $t$ be a sequence in $D$, and the utility-array of $t$ is denoted as $t.ua$. The $t$-projected utility-array, denoted as $(D.ua)|_t$, is the collection of \textit{suffix} of arrays in $D.ua$ w.r.t. prefix $t$. 
\end{definition}

\begin{table}[!htbp] 
	\setlength{\abovecaptionskip}{0pt}
	\setlength{\belowcaptionskip}{0pt}
	\caption{The projected utility-array of $<$$[a][c]$$>$ in $ S_{1} $.}
	\label{table:projectUA}
	\centering
	\begin{tabular}{|c|c|c|c|c|c|c|}
		\hline
		&  \textit{\textbf{eid}}  & \textit{\textbf{item}}  &  \textit{\textbf{u}}  &  \textit{\textbf{ru}}  &  \textit{\textbf{next\_pos}}  &  \textit{\textbf{next\_eid}} \\ \hline \hline
		\textit{array$_4$} & 3   & $b$  & \$20  &  \$15 &  -  & 6 \\ \hline
		\textit{array$_5$} & 3   & $f$  & \$3  &  \$12 &  -  & 6 \\ \hline
		\textit{array$_6$} & 4   & $a$  & \$6  &  \$6 &  -  & - \\ \hline
		\textit{array$_7$} & 4   & $e$  & \$6  &  \$0 &  -  & - \\ \hline
		
	\end{tabular}
\end{table}

For example, the projected utility-array of $<$$[a]$ $[c]$$>$ in $ S_{1} $ is shown in Table \ref{table:projectUA}. Similarly, the other utility-arrays of $<$$[a]$ $[c]$$>$ can be projected in other sequences, for example, $ S_{2} $ and $ S_{4} $. As an accurate representation, utility-array provides provably equivalent decomposition as projected database from the original sequence data, but it requires much less memory space. It proceeds by dividing the initial utility-arrays into smaller ones projected on the subsequences that were obtained so far, and only their corresponding suffixes are kept. The number of transactions/sequences in the projected utility-array is less than original database. This can substantially reduce the cost of the projection operation when the projected utility-arrays can be held in the main memory. By combining sequences from a series of projected utility-arrays, all the HUSPs and the candidates can be acquired.

It is important to notice that, instead of constructing the projected sub-database that only contain the updated sequences, a set of the projected compact utility-array of each sequence is only constructed $s \in D$. In other words, only the projected utility-arrays in each projection process are constructed and then scanned for constructing the next updated ones. Different from previous studies \cite{wang2016efficiently,yin2012uspan} that require scanning the projected sub-database to construct the data structure (e.g., utility-matrix \cite{yin2012uspan} and utility-chain \cite{wang2016efficiently}), the proposed ProUM algorithm does not need to construct and scan the projected sub-database.

Based on the designed utility-array and its construction process, the following desirable properties of the utility-array can be obtained: (1) The obtained information from utility-array is exact. Since the utility-array of ($l$+1)-sequence is constructed based on the built utility-array of $l$-sequences, it is parameter-free and contains the complete information. (2) It is space efficient because it requires an inconsequential space overhead to construct and project a series of utility-arrays, which allows massive sequences to be processed in main memory (for most data mining, disk is death). (3) It has simplicity and intuitiveness because it can be constructed in deterministic time and regarded as the representation of quantitative sequences.

\subsection{Proposed Upper Bound and Pruning Strategies}

It is known that the complete search space of SPM or HUSPM is much more difficult than FIM. For HUSPM, it has 2$^{m \times n}$ possible candidates in total, where $m$ is the total number of all the possible distinct items in the database, and $n$ is the number of elements in the longest sequence. However, in HUSP mining, the downward closure property (e.g., the Apriori property \cite{agrawal1995mining,srikant1996mining}) does not hold for the utility of sequence patterns. Because \textit{SWU} has the downward closure property, the current algorithms (e.g., USpan \cite{yin2012uspan} and HuspExt \cite{alkan2015crom}) adopt \textit{SWU} to prune the search space. However, they usually suffer from the problem of an exponential number of candidate sequences since \textit{SWU} is a loose upper bound to over-estimate the true utility of a sequence. 

An optimization with a new upper bound is proposed below to further improve ProUM's efficiency. The search space of ProUM can be systematically explored by utilizing the presented pruning strategies.

\begin{definition}
	\rm The sequence-weighted utilization (\textit{SWU}) \cite{yin2012uspan} of a sequence $ t $ in a quantitative sequential database $ D $ is denoted as  \textit{SWU}$(t) $ and defined as follows:
	\begin{equation}
		SWU(t) = \sum_{t \subseteq s \wedge s\in D}u(s).
	\end{equation}
\end{definition}

For example, in Table \ref{table:db}, \textit{SWU}($<$\textit{a}$>$) = $ u(S_{1}) $ + $ u(S_{2}) $ + $ u(S_{3}) $ + $ u(S_{4}) $ + $ u(S_{5}) $  = \$71 + \$69 + \$38 + \$63 + \$52 = \$293, and \textit{SWU}($<$[$a$] [$c$]$>$) = $ u(S_{1})$ + $ u(S_{2})$ + $ u(S_{4})$ = \$71 + \$69 + \$63 = \$203.

\begin{theorem}(global downward closure property \cite{yin2012uspan}) 
	\label{SWU_theorem}
	\rm Given a quantitative sequential database $ D $ and two sequences $ t $ and $ t' $, if $ t\subseteq t' $, then:
	\begin{equation}
		SWU(t')\leq SWU(t).
	\end{equation}	
\end{theorem}

\begin{theorem}
	\label{u(t)_theorem}
	\rm  Given a quantitative sequential database $ D $ and a sequence $ t $, it can be obtained that:
	\begin{equation}
		u(t)\leq SWU(t).
	\end{equation}	
\end{theorem}

The proof for Theorem \ref{SWU_theorem} and Theorem \ref{u(t)_theorem} can further referred to in \cite{alkan2015crom,yin2012uspan}. To improve the performance of utility mining, the USpan algorithm \cite{yin2012uspan} introduces an upper bound based on the remaining utility concept. Details are introduced below.

\begin{definition}(first match)
	\rm  Given two $ q $-sequences $ s $ and $ s' $, if $ s\subseteq s' $, then the extension of $ s $ in $ s' $ is said to be the rest of $ s' $ after $ s $, and is denoted as $<$$ s' $-$ s $$>_{rest}$. 
	Given a sequence $ t $ and a $ q $-sequence $ s $, if $ t \sim s_{k} \wedge s_{k} \subseteq s $ $ (t \subseteq s) $, the rest of $ t $ in $ s $ is the rest part of $ s $ after $ s_{k} $, which is denoted as $<$$s$-$t$$>_{rest}$, where $ s_{k} $ is the first match of $ t $ in $ s $. 
\end{definition}

As an example, consider the sequence $ t $ = $<$[\textit{a}], [\textit{b}]$>$, \textit{q}-sequences $ s $ = $<$[\textit{a}:4], [\textit{b}:2]$>$, and $ S_{3} $ in Table \ref{table:db}. The remaining part of $ s $ in $ S_{3} $ is $<$$S_{3}$ - $s$$>$$_{rest}$ = $<$[(\textit{e}:1)]$>$, and thus it is unique. However, two remaining parts of $ t $ exist in $ S_{3} $ since it has two matches of $ t $ in $ S_{3} $, and the first one is $<$[\textit{a}:4], [\textit{b}:4]$>$. Based on our definition, $<$$S_{1}$ - $t$$>$$_{rest}$ = $<$[(\textit{f}:5)], [(\textit{a}:1) (\textit{b}:2) (\textit{e}:1)]$>$.

\begin{definition} (sequence-projected utilization, SPU \cite{yin2012uspan})
	\rm  The sequence-projected utilization (\textit{SPU}) of a sequence $t$ in a sequence database $D$ is denoted as \textit{SPU}($t$) and defined as follows: 
	\begin{equation}
	    SPU(t) = \sum_{i \in s' \wedge s' \subseteq s \wedge s \in D}(u_{rest}(i,s) + u_{p}(t,s)),
	\end{equation} 
	 where $i$ is the pivot of $t$ in $s$, and $u_{rest}(i, s)$ is referred to the remaining utility at $q$-item $i$ (exclusive) in $q$-sequence $s$, such as $u_{rest}(i,s)$ =  $\sum_{i'\in s \wedge i \prec i'}u(i')$. $u_{p}(t,s)$ is the utility of $t$ in position pivot (\textit{p}) in $s$. Note that the pivot is the first place where the $q$-subsequences match $t$.
\end{definition}

Thus, the \textit{SPU} value of sequence $t$ is the sum of the remaining utilities and utilities of the far left subsequences that match $t$. However, this is not a true upper bound on utility. When using \textit{SPU} to prune the search space in the LQS-tree with DFS strategy, it may miss some of the real HUSPs. Other reports of this 
serious problem can be referred to in \cite{truong2019survey}. Therefore, we propose a real upper-bound on utility for mining HUSPs and the details are given below.

\begin{definition} (sequence extension utility of $t$ in $s$)
	\rm   The sequence extension utility (\textit{SEU}) is used to present the maximum utility of the possible extensions that based on the prefix $t$. Let \textit{SEU}$(t, s)$ denote the \textit{SEU} of a sequence $t$ in $s$, and it indicates how much of the sequence's overall utility remains to be extended/concatenated. It is defined as follows: 
	\begin{equation}
		SEU(t, s) = u_{rest}(i,s) + u(t,s),
	\end{equation}	
	  where $u_{rest}(i,s)$ is the remaining utility of $i$ in $s$, $i$ is the pivot of $t$ in $s$, w.r.t. is the first occurring position of $s'\sim t$, and $u(t,s)$ is the maximum utility of $t$ in $s$.
\end{definition} 

Based on the definition of $<$$s - t$$>$$_{rest}$ (cf. Definition 24), it can be seen that $u_{rest}(i,s)$ is equal to $<$$s - t$$>$$_{rest}$. For consistency, $<$$s - t$$>$$_{rest}$ is used in the following contents.

\begin{definition}(sequence extension utility $t$ in $D$)
	\rm The overall sequence extension utility of a sequence $ t $ in a quantitative sequential database $ D $ is denoted as \textit{SEU}$(t) $ and defined as follows:
	\begin{equation}
		SEU(t) = \sum_{t \subseteq s \wedge s\in D} (u(t,s) + u(<s - t>_{rest})),
	\end{equation}
	where $u(t,s)$ is the maximum utility value of $t$ in $s$.
\end{definition}

It should be noted that $u(t,s)$ is the maximum utility of $t$ in $s$ (also can be referred to Definition 13) while the $u_{p}(t,s)$ cannot guarantee the maximum utility of $t$ in $s$. \textit{SEU} is different from \textit{SPU}. Intuitively, in each $q$-sequence $s$ that $t \subseteq s$, \textit{SEU} contains less utility values than \textit{SWU}.

Note that \textit{u}($<$\textit{s} - \textit{t}$>_{rest}$) can be obtained from the constructed utility-array of $ t $ in $ s $, which contains the remaining utility of $ t $ in each position. For example, in the sequence $ t $= $<$[\textit{a}], [\textit{b}]$>$ in Table \ref{table:db}, \textit{SEU}$(t) $ = $ u(t, S_{1})$ + \textit{u}($<$$S_{1}$ - $t$$>$$_{rest}$) + $ u(t, S_{2})$ + \textit{u}($<$$S_{2}$ - $t$$>$$_{rest}$) + $ u(t, S_{3})$ + \textit{u}($<$$S_{3}$ - $t$$>$$_{rest}$) = (\$18 + \$15) + (\$23 + \$16) + (\$20 + \$18) = \$110.

\begin{theorem}(local downward closure property)
	\rm Given a quantitative sequential database $ D $ and two sequences $ t $ and $ t' $, if $ t\subseteq t' $, it can be obtained that: 
	\begin{equation}
		SEU(t')\leq SEU(t).
	\end{equation}		
\end{theorem}

\textbf{Proof}.  Suppose that $ s_{q'} $ is a \textit{q}-sequence that satisfies $ u(s_{q'}) $ = $ u(t', s) $, where $ t' \sim s_{q'} \wedge s_{q'} \subseteq s \wedge s \in D $. The sequence $ t' $ can be divided into two parts as the prefix $ t $ and the extension $ e $ such that $ t $ + $ e $  = $ t' $. 
	Correspondingly, the sequence $ s_{q'} $ can also be divided into two parts as the prefix $ s_{q'_{t}} $ matching $ t $ and the extension $ s_{q'_{e}} $ matching $ e $ such that $ s_{q'_{t}} $ + $ s_{q'_{e}} $ = $ s_{q'} $. 
	Then, we have:
	\begin{align*}
	SEU(t', s)  &=  u(t', s) + u(<s - t'>_{rest})\\
	&=  u(s_{q'_{t}}) + u(s_{q'_{e}}) + u(<s - t'>_{rest})\\
	&\leq u(t, s) + u(s_{q'_{e}}) +  u(<s - t'>_{rest})\\
	&\leq u(t, s) + u(<s - t>_{rest})\\
	& = SEU(t, s).
	\end{align*}

	Thus, $SEU(t', s)$ $\leq SEU(t, s)$. According to $ t\subseteq t' $, it is obtained that the set of sequences where $ t' \subseteq s $ is a subset of that of $ t \subseteq s $. Therefore, $\displaystyle SEU(t')$ = $\sum_{t' \subseteq s \wedge s\in D} \{u(t',s)$ + $u($$<$$s - t'$$>$$_{rest})\}$ $ \leq \sum_{t' \subseteq s \wedge s\in D} \{u(t,s)$ + $u($$<$$s - t$$>$$_{rest})\} \leq \sum_{t \subseteq s \wedge s\in D} \{u(t,s)$ + $u($$<$$s - t$$>$$_{rest})\}$ = $SEU(t)$. So far, this theorem holds.

\begin{theorem}
	\label{SEU_upperbound}
	\rm The \textit{SEU} value of a sequence $ t $ is an upper bound on the utility of this sequence in a quantitative sequential database $ D $. It always has the following relationship:
	\begin{equation}
		u(t)\leq SEU(t).
	\end{equation}	
\end{theorem}

\textbf{Proof}. We have that $\displaystyle u(t)$ = $\sum_{t \subseteq s \wedge s\in D} \{u(t,s)\} $ $ \leq \sum_{t \subseteq s \wedge s\in D} \{u(t,s)$ + $u($$<$$s - t$$>$$_{rest})\}$ = $SEU(t)$.

\begin{definition} (\textbf{promising HUSP}) 
	\rm  A sequence $t$ in $D$ is called a \textit{promising} high-utility sequential pattern  iff: 1) if the node for $t$ is an \textit{I}-\textit{Concatenation} node and satisfies \textit{SWU}$(t) \geq \delta \times u(D)$ or \textit{SEU}$(t) \geq \delta \times u(D)$; and 2) if the node for $t$ is an \textit{S}-\textit{Concatenation} node and satisfies \textit{SWU}$(t) \geq \delta \times  u(D)$ or \textit{SEU}$(t) \geq \delta \times  u(D)$; otherwise, this sequence/node is called an invalid or \textit{unpromising} pattern.

\end{definition}

Based on Theorem \ref{SEU_upperbound}, if the upper bound  \textit{SEU} is less than $\delta $ $\times $ $ u(D)$, then ProUM can be directly stopped from going deeper and the search procedure can be backtracked. The \textit{SEU} value of a sequence is an upper bound on the utilities of its extension (the part of its super-sequences). It should be noted that the \textit{SEU} has the local downward closure property but not the global downward closure property. The reason for this is that some super-sequences of a node/sequence $t$ are not in the subtree of $t$, and they may be the promising HUSPs even though $t$'s \textit{SEU} value is less then $\delta $ $ \times $ $ u(D)$.  \textit{SEU} can be used to effectively prune the search space in finding HUSPs. In general, \textit{SEU} is tighter than the previous upper bound \textit{SWU}. Note that for any non-root node $N$ in the LQS-tree, the \textit{SEU} can be quickly obtained as an upper bound of all the nodes in the subtree rooted at node $N$.

\begin{strategy} 
   \rm  \textbf{(Pruning of the unpromising one-$q$-sequences by \textit{SWU}, called the PUO strategy)}: Let $t$ be the sequence represented by a node $N$ in the LQS-tree,  $t'$ be represented as a child node of $N$, and $\delta$ be the minimum utility threshold. If \textit{SWU}$(t) \geq \delta $ $ \times$  $u(D)$, then ProUM can be stopped from exploring node $N$. The reason for this is that the sequence $t'$ is always a super-sequence of $t$. Hence, $ u(t') \leq $ \textit{SWU}$(t') \leq $ \textit{SWU}$(t) < \delta $ $\times$ $u(D)$. The upper bound \textit{SWU} has the global downward closure property, and thus any super-sequence $t'$ and its extensions cannot be a desired HUSP. Note that the PUO is a global pruning strategy.
\end{strategy}

\begin{strategy} 
	\rm  \textbf{(Pruning of the unpromising $k$-$q$-sequences by \textit{SEU}, called the PUK strategy)}: The upper bound \textit{SEU} of a sequence $t$ can be utilized to prune the unpromising $k$-$q$-sequence in its subtree at an early stage when traversing the LQS-tree with the DFS strategy. Thus, if \textit{SEU}$(t) < \delta \times  u(D)$, then the generation of the utility-arrays of its \textit{I}-\textit{Concatenation} and \textit{S}-\textit{Concatenation} can be stopped, and traversing all the subtrees from $t$ can be stopped. This is because the utility of $t$ and any of $t$'s offspring would not more than \textit{SEU}$(t)$. Note that the PUK is a local pruning strategy that can be used in the depth pruning in LQS-tree, for example, $I$-extension pruning and $S$-extension pruning.
\end{strategy} 

For example, to avoid constructing the utility-array of the unpromising ($k$+1)-sequences, the PUK strategy can be applied when scanning the projected sub-databases for $k$-sequences. This filter operation can reduce both the execution time and memory cost. In summary, the PUO strategy can be used for both width pruning and depth pruning, but it is only used for width pruning in the proposed ProUM algorithm; ProUM utilizes the more powerful PUK strategy for depth pruning. The former also affects the performance of the later. With the PUK strategy, ProUM can easily be stopped from going deeper and the search procedure can be backtracked.

\subsection{Proposed ProUM Algorithm}

The details of LQS-tree, utility-array, the projection mechanism, and the pruning strategies with \textit{SWU} and \textit{SEU} have been introduced as far. To summarize, the pseudocode of main procedure of ProUM is shown in Algorithm \ref{AlgorithmProUM}. A quantitative sequence database $D$, a profit-table $ptable$, and a minimum utility threshold $\delta$ are contained in the input for ProUM; the output includes all the high-utility sequential patterns (HUSPs). Without loss of generality, it is assumed that the proposed ProUM traverses the LQS-tree using the Depth-First-Search (DFS) strategy. By deleting the 1-sequences that \textit{SWU}$(t) < \delta$ $\times$ $u(D)$ (Line 3), it first scans the original database once to obtain the \textit{SWU} value of each 1-sequence $t \in D$ (Line 2), and then the revised database $D'$ is obtained. Then, the revised database $D'$ is scanned once to construct the initial utility-arrays for all the sequences in $D'$ (Line 4). After that, ProUM recursively projects a series of sub-utility-arrays based on the prefix sequences (Line 5) by traversing the LQS-tree with DFS strategy.

\renewcommand{\algorithmicrequire}{\textbf{Input:}}
\renewcommand{\algorithmicensure}{\textbf{Output:}}
\begin{algorithm}
	\caption{The ProUM algorithm}
	\label{AlgorithmProUM}

	\begin{algorithmic}[1]		
		\REQUIRE \textit{D}; \textit{ptable}; $\delta$.
		\ENSURE \textit{HUSPs}: the complete set of high-utility sequential patterns.      
		
		\STATE initialize $D.ua$ = $ \emptyset$;\
		\STATE scan the original database once to get the \textit{SWU} value of each 1-sequence;\
		\STATE get the revised database $D'$, by deleting the 1-sequences that \textit{SWU}$(t) < \delta$ $\times$ $ u(D)$ (\underline{the PUO strategy});\
		\STATE scan the revised database $D'$ once to construct the initial utility-arrays $D.ua$ for all sequences in $D'$;\
		\STATE \textbf{call Project-Search}$\boldmath{(\emptyset, I^*, D.ua, \delta)}$.\		
		\STATE \textbf{return} \textit{HUSPs}
	\end{algorithmic}
\end{algorithm}

\renewcommand{\algorithmicrequire}{\textbf{Input:}}
\renewcommand{\algorithmicensure}{\textbf{Output:}}
\begin{algorithm}
	\caption{The Project-Search procedure}
	\label{AlgorithmProjectSpan}
	
	\begin{algorithmic}[1]		
		\REQUIRE \textit{t}: a sequence as prefix; $(D.ua)|_t$: the projected utility-array of $t$; $\delta$: the minimum utility threshold.
		\ENSURE \textit{HUSPs}: the set of high-utility sequential patterns with prefix $t$.      
		
		\STATE initialize $iItem$ = $ \emptyset$ and $sItem$ = $ \emptyset$;\
		\STATE scan the projected utility-array $(D.ua)|_t$ once to: \\
		1) put $I$-Concatenation items of $t$ into \textit{iItem}; \\
		2) put $S$-Concatenation items of $t$ into \textit{sItem}; \\
		3) calculate the \textit{SEU} values of these items form $(D.ua)|_t$;
		
		\STATE remove unpromising items $i_j \in iItem$ that have  \textit{SEU}$(i_j) < \delta$ $\times $ $ u(D)$  (\underline{the PUK strategy});\
		\STATE remove unpromising items $i_j \in sItem$ that have  \textit{SEU}$(i_j) < \delta$ $\times $ $ u(D)$  (\underline{the PUK strategy});\
		
		\FOR {each item $ i \in iItem$}
			\STATE $ t' \leftarrow $ $I$-\textit{Concatenation}$(t, i) $;\
			\STATE construct the projected utility-array $(D.ua)|_{t'}$;\
			\IF { \textit{SEU}$(t') \geq \delta$ $\times$ $u(D)$}
			   \IF {$ u(t') \geq \delta$ $\times$ $u(D)$}
			   		\STATE  output $t'$ into \textit{HUSPs};\
			   \ENDIF
			   \STATE \textbf{call Project-Search}$(t', (D.ua)|_{t'}, \delta)$.\ 
			\ENDIF						    
		\ENDFOR

		\FOR {each item $ i \in sItem$}
			\STATE $ t' \leftarrow $ $S$-\textit{Concatenation}$(t, i) $;\
			\STATE construct the projected utility-array $(D.ua)|_{t'}$;\
			\IF { \textit{SEU}$(t') \geq \delta$ $\times$ $u(D)$}
				\IF {$ u(t') \geq \delta$ $\times$ $u(D)$}
					\STATE  output $t'$ into \textit{HUSPs};\
				\ENDIF
				\STATE \textbf{call Project-Search}$(t', (D.ua)|_{t'}, \delta)$.\ 
			\ENDIF						
		\ENDFOR 		
		
		\STATE \textbf{return} \textit{HUSPs}
	\end{algorithmic}
\end{algorithm}

The details of the projection and searching procedure are presented in Algorithm \ref{AlgorithmProjectSpan}. When visiting a node/sequence $t$, ProUM first initializes two sets, $iItem$ = $ \emptyset$ and $sItem$ = $ \emptyset$ (Line 1). Then, to obtain the promising items for $I$-\textit{Concatenation} and $S$-\textit{Concatenation} (Line 2), it scans the projected utility-array $(D.ua)|_t$ once. Note that the \textit{SEU} value of each item is calculated simultaneously during the utility-array scanning (Line 2). After obtaining the updated two sets of $iItem$ and $sItem$, ProUM removes unpromising items that have \textit{SEU}$(i_j) < \delta$ $\times $ $ u(D)$ in $iItem$ and $sItem$, respectively (Lines 3 to 4, the PUK strategy). Next, all these items in $iItem$ and $sItem$ may be used to generate the promising extensions $t'$ as descendant of $t$. The process of items in $iItem$ is shown in Lines 5 to 14. For a new extension $t'$ whose prefix is $t$ (Line 6), it first constructs the projected sub-utility-array $(D.ua)|_t'$ based on the previous utility-array $(D.ua)|_t$ (Line 7). At the same time, the \textit{SEU} value of this $I$-\textit{Concatenation} can be calculated. Then, ProUM checks this $t'$ whether is able to be the extension as descendant of $t$. The PUK strategy is used (Line 8, using \textit{SEU} upper bound). If its \textit{SEU} $< \delta$ $\times $ $ u(D)$, then ProUM backtracks to the parent of $t$; otherwise, ProUM continues to check the overall utility of this extension $t'$ and outputs $t'$ as a final HUSP if  $t'$ satisfies $\delta$ $\times $ $ u(D)$ (Lines 9 to 11). Additionally, ProUM calls the \textbf{Project-Search} procedure to begin the next projection and search with respect to prefix $t'$ (Line 12). Finally, ProUM recursively explores the other extension nodes in $iItem$ (Line 5) in a similar manner. Similarly, ProUM performs the above procedure to handle each $S$-extension item in \textit{sItem} (Lines 15 to 24).

\textbf{Implementation details}. During the $t$-projected utility-array scan with respect to $s$, to calculate the \textit{SEU}$(t, s)$ value, an intuitive method is to scan all arrays in the utility-array of $s$. ProUM has an efficient implementation. The \textit{SEU} value of the sequence is obtained simultaneously when constructing the utility-array of a sequence/transaction. Thus, \textit{SEU} of $s$ is added into the utility-array of $s$. By also storing this \textit{SEU} value, we can avoid scanning all the elements in the utility-array for calculating the upper bound \textit{SEU} of $t$ in $D$. Thus, for a given sequence $t$, its \textit{SEU} value can be quickly obtained since it can be accumulated from the stored \textit{SEU} values by a set of sequences/transactions w.r.t. \textit{sid}.

\begin{table*}[h]
	
	\caption{Dataset features}
	\label{tab:datasets}
	
	\centerline{
		\begin{tabular}{| c | l | c | c | c | c | c | c | l |}
			\hline
			& \textbf{Dataset} & \textbf{\#$|D|$} & \textbf{\#$|I|$} & \textbf{avg(\#S)} & \textbf{max(\#S)} & \textbf{avg(\#Seq)} & \textbf{ave(\#Ele)} & \textbf{description} \\
			\hline \hline
			\multirow{6}{*}{1.} 
			& Sign & 730  & 267  & 52 & 94 &  51.99 & 1.0 & language utterance  \\
			\cline{2-9}  
			& Bible & 36,369 & 13,905 & 21.64 & 100 & 17.85  & 1.0 & text \\
			\cline{2-9}  			
			& SynDataset-160k & 159,501 & 7,609 & 6.19 & 20 & 26.64 & 4.32 & synthetic sequences\\
			\cline{2-9} 			
			& Kosarak10k & 10,000 & 10,094 & 8.14 & 608 & 8.14 & 1.0 & web click stream\\
			\cline{2-9}  
			& Leviathan & 5,834 & 9,025 & 33.81 & 100 & 26.34  & 1.0 & text\\
			\cline{2-9}  			
			& yoochoose-buys & 234,300 & 16,004 & 1.13 & 21 & 2.11 & 1.97 & purchase data\\
			\hline			
			\multirow{5}{*}{2.} 
			& C8S6T4I3D$|$X$|$K (10k) & 10,000 & 7,312 & 6.22 & 18 & 26.99 & 4.35 & synthetic dataset\\
			\cline{2-9}  
			& C8S6T4I3D$|$X$|$K (80k) & 79,718 & 7,584 & 6.19 & 18 & 26.69 & 4.32 & synthetic dataset\\
			\cline{2-9}  
			& C8S6T4I3D$|$X$|$K (160k) & 159,501 & 7,609 & 6.19 & 20 & 26.64 & 4.32 & synthetic dataset\\
			\cline{2-9} 
			& C8S6T4I3D$|$X$|$K (240k) & 239,211 & 7,617 & 6.19 & 20 & 26.66 & 4.32 & synthetic dataset\\
			\cline{2-9}  
			& C8S6T4I3D$|$X$|$K (320k) & 318,889 & 7,620 & 6.19 & 20 & 26.64 & 4.32 & synthetic dataset\\
			& C8S6T4I3D$|$X$|$K (400k) & 398,716 & 7,621 & 6.18 & 20 & 26.64 & 4.32 & synthetic dataset\\
			\hline
	\end{tabular}}
\end{table*}

\begin{figure*}[htbp]
	\centering 
	\includegraphics[trim=20 5 10 0,clip,scale=0.56]{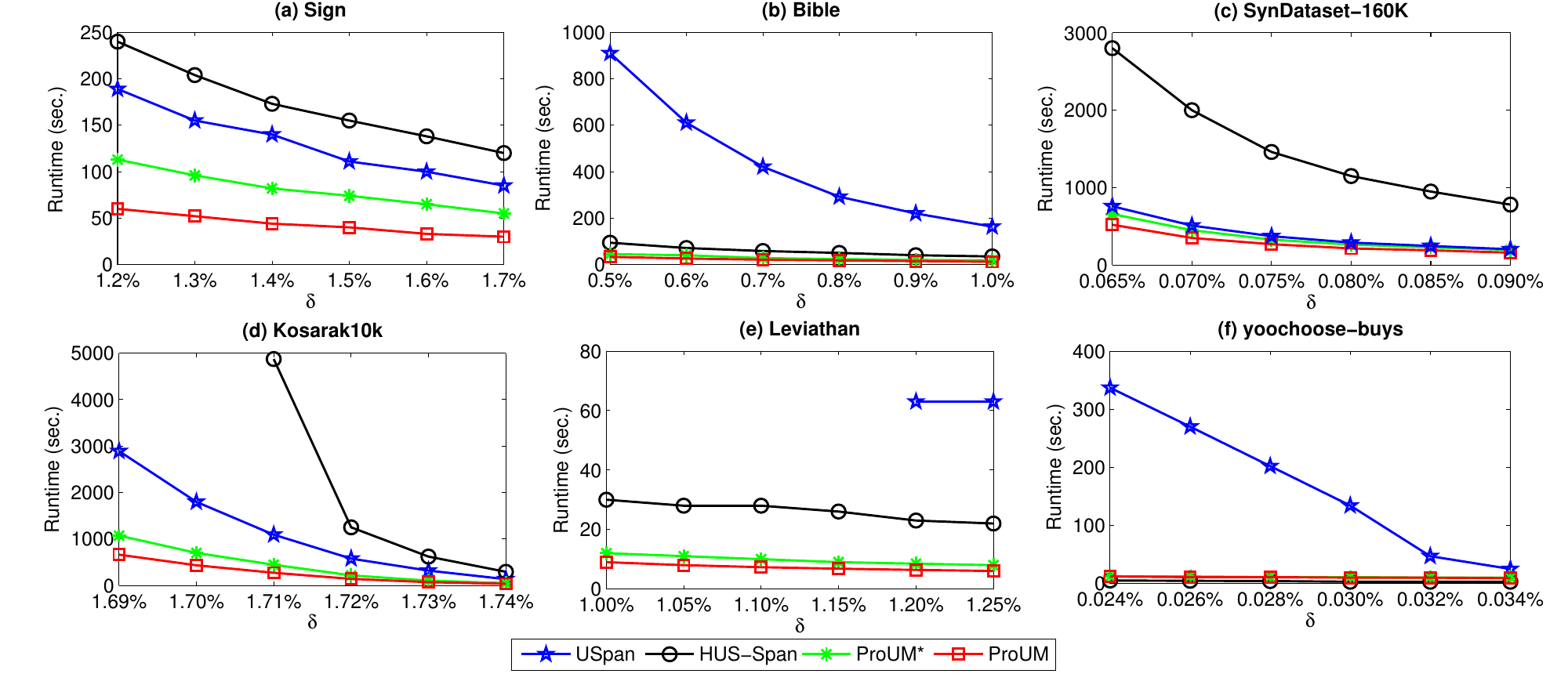}
	\captionsetup{justification=centering}
	\caption{Runtime by varying $\delta$}
	\label{fig:Runtime}	
\end{figure*}

\begin{table*}[htb]
	
	\fontsize{6.8pt}{9pt}\selectfont
	\centering
	\caption{Number of patterns (candidates and final results) under various $ \delta $ values}
	\label{table:pattern}
	\begin{tabular}{|cc|llllll|}
		\hline\hline
		\multirow{2}*{\textbf{}}&
		\multirow{2}*{\textbf{}}
		&\multicolumn{6}{|c|}{\textbf{\# of patterns under various $ \delta $ values}}\\
		\cline{3-8}
		&& \textit{$ \delta $}$_1$ & \textit{$ \delta $}$_2$ & \textit{$ \delta $}$_3$ &  \textit{$ \delta $}$_4$ &  \textit{$ \delta $}$_5$  &  \textit{$ \delta $}$_6$ \\ \hline

		&  \textbf{\#\textit{P}1} & 6,598,217 &	5,265,824 &	4,250,363 &	3,490,867 &	2,886,277 &	2,418,799  \\
		&  \textbf{\#\textit{P}2} & 6,598,217 &		5,265,825 &		4,250,363 &		3,490,869 &		2,886,278 &		2,418,799 \\
		($a$)  Sign &  \textbf{\#\textit{P}3} & 24,586,054 & 	19,345,657 & 	15,424,452 & 	12,536,127 & 	10,269,881 & 	8,534,821   \\
		& \textbf{\#\textit{P}4} & 6,598,215 & 	5,265,822 & 	4,250,359 & 	3,490,865 & 	2,886,274 & 	2,418,798  	\\
		&  \textbf{\#\textit{HUSPs}} & \textbf{78,336} &	\textbf{56,395}	 &   \textbf{41,151} &	\textbf{30,440} & 	\textbf{22,702} &	  \textbf{17,274}  \\
		\hline

		&  \textbf{\#\textit{P}1} & 92,563 &	59,041 &	40,766 &	29,183 &	21,227 &	16,488 \\
		&  \textbf{\#\textit{P}2} & 95,012 & 	60,588 & 	41,786 & 	29,940 & 	21,746 & 	16,887 \\
		($b$)   Bible &  \textbf{\#\textit{P}3} & 262,465 & 	163,564 & 	109,449 & 	76,979 & 	56,443 & 	42,205	 \\
		& \textbf{\#\textit{P}4} & 100,706 & 	64,091 & 	43,983 & 	31,630 & 	23,486 & 	17,831	\\
		&  \textbf{\#\textit{HUSPs}}  & \textbf{2,760} &	\textbf{1,714} &	\textbf{1,124} &	\textbf{764} &	\textbf{553} &	\textbf{411}	 \\
		\hline

		&  \textbf{\#\textit{P}1} & 8,751,355 &	5,357,157 &	3,313,183 &	2,122,646 &	1,394,602 &	948,156 \\
		&  \textbf{\#\textit{P}2} & 8,752,654 &	5,357,847 &	3,313,681 &	2,123,007 &	1,394,913 &	948,397 \\
		($c$)   SynDataset-160K &  \textbf{\#\textit{P}3} & 23,917,337 &	13,187,663 &	8,101,205 &	5,412,965 &	3,735,188 &	2,645,849	 \\
		& \textbf{\#\textit{P}4} & 8,753,634 &	5,358,865 &	3,314,487 &	2,123,613 &	1,395,359 &	948,520	\\
		&  \textbf{\#\textit{HUSPs}}  & \textbf{58,710} &	\textbf{17,903} &	\textbf{4,794} &	\textbf{1,344} &	\textbf{394} &	\textbf{172}	 \\
		\hline

		&  \textbf{\#\textit{P}1} & 124,833,772 &	82,478,688 &	51,535,423 &	24,542,377 &	12,100,103 &	5,524,463 \\
		&  \textbf{\#\textit{P}2} & - &	- &	51,535,979 &	24,542,940 &	12,100,669 &	5,525,033 \\
		($d$) Kosarak10k &  \textbf{\#\textit{P}3}  & 478,850,673 &	321,429,317 &	204,661,580 &	100,469,791 &	42,141,209 &	20,427,300	 \\
		& \textbf{\#\textit{P}4} & 124,833,676 &	82,478,593 &	51,535,330 &	24,542,295 &	12,100,024 &	5,524,390	\\
		&  \textbf{\#\textit{HUSPs}}  & \textbf{23} &	\textbf{22} &	\textbf{22} &	\textbf{22} &	\textbf{22} &	\textbf{21}	 \\
		\hline

		&  \textbf{\#\textit{P}1} & -  &  -  &  -  &  -  &  46,193 & 	41,610 \\
		&  \textbf{\#\textit{P}2} & 76,549 &	68,058 &	60,557 &	54,084 &	48,162 &	43,361 \\
		($e$) Leviathan &  \textbf{\#\textit{P}3}  & 205,381 &	181,067 &	159,208 &	140,946 &	125,177 &	111,605	 \\
		& \textbf{\#\textit{P}4} & 82,625 &	73,315 &	65,076 &	58,140 &	52,181 &	47,031	\\
		&  \textbf{\#\textit{HUSPs}}  & \textbf{1,802} &	\textbf{1,520} &	\textbf{1,322} &	\textbf{1,152} &	\textbf{9,96} &	\textbf{869}	 \\
		\hline

		&  \textbf{\#\textit{P}1} & 325,473  &  304,534  &  278,131  &	250,138  &	201,440  &	158,219  \\
		&  \textbf{\#\textit{P}2} & 312,780 &	304,737 &	278,318 &	250,307 &	201,620	 & 158,409  \\
		($f$)  yoochoose-buys &  \textbf{\#\textit{P}3} & 314,346 &	305,226 &	280,022 &	251,896 &	203,867 &	159,858	 \\
		& \textbf{\#\textit{P}4} & 313,724 &	304,642 &	279,481 &	251,434 &	203,422 &	159,463	\\
		&  \textbf{\#\textit{HUSPs}}  & \textbf{317,682}  &	\textbf{296,890}  &	\textbf{273,926}  &	\textbf{238,273}  &	\textbf{191,203}  &	\textbf{146,761} \\
		
		\hline\hline
	\end{tabular}
\end{table*}

\section{Experiments} \label{sec:experiments}

Several experiments were conducted to demonstrate the effectiveness and efficiency of the proposed projection-based utility mining ProUM algorithm. 

\textbf{Evaluation metric}. In general, the evaluation metric of comparison for utility-oriented sequential mining algorithms consists of effectiveness analysis with derived patterns, efficiency analysis with execution time and memory consumption, and scalability evaluation. In the following subsections, we use these metrics to evaluate the performance of ProUM.

\textbf{Compared baselines}. For efficiency analysis, USpan \cite{yin2012uspan} (replacing the original \textit{SPU} by \textit{SEU}) and the state-of-the-art HUS-Span \cite{wang2016efficiently} algorithm were selected as the baselines. It should be noted that the CRoM that was used in HuspExt \cite{alkan2015crom} is not a true upper bound, and the discovered results by HuspExt are not complete. Therefore, HuspExt is not compared in the following experiments. Two variants of ProUM (respectively denoted as ProUM$^*$, and ProUM) were compared to evaluate the effect of the proposed pruning strategies. ProUM is a hybrid optimization algorithm, as shown in Algorithm \ref{AlgorithmProUM} and Algorithm \ref{AlgorithmProjectSpan}. In addition, the difference between ProUM and ProUM$^*$ is that in Algorithm \ref{AlgorithmProjectSpan} Lines 3 to 4, ProUM$^*$ adopts the PUO strategy to filter the unpromising items in \textit{iItem} and \textit{sItem}.

\subsection{Data Description and Experimental Configuration}

\textbf{Datasets}. For the performance tests, a total of seven datasets were chosen for the different characteristics they displayed. The goal was to show the efficiency of the developed algorithm in a wide range of situations. The chosen datasets and their characteristics are displayed in Table \ref{tab:datasets}. Note that \#$|D|$ is the number of sequences, \#$|I|$ is the number of different symbols/items in the dataset, \#S is the length of a sequence $s$, \#Seq is the number of elements per sequence, and \#Ele is the average number of items per element/itemset. SynDataset-160K is a synthetic sequential dataset generated by IBM Quest Dataset Generator \cite{agrawal1994dataset}. The original yoochoose-buys\footnote{\url{https://recsys.acm.org/recsys15/challenge/}} dataset contains the quantity and unit profit of each object/item while other datasets\footnote{\url{http://www.philippe-fournier-viger.com/spmf/index.php}} do not contain the quantity and unit profit. Therefore, we adopted a simulation method that is widely used in previous studies \cite{lin2017fdhup,liu2012mining,tseng2013efficient} to generate the quantitative and profit information for each object/item in datasets except for yoochoose-buys.

\textbf{Experimental configuration}. All the compared algorithms in the experiments were implemented in Java language. Note that the original USpan algorithm with \textit{SPU} upper bound may cause incomplete mining results. Thus, the USpan code used here is a revised and optimized version. Furthermore, \textit{SPU} was replaced with \textit{SEU} in USpan so that it could discover the complete HUSPs.  All the experiments were performed on a personal ThinkPad T470p computer with an Intel(R) Core(TM) i7-7700HQ CPU @ 2.80 GHz 2.81 GHz processor, 32 GB of RAM, and with 64-bit Microsoft Windows 10 operating system.

\subsection{Efficiency Analytics}

As previously mentioned, a good high-utility sequence mining method should be efficient and able to scale well to handle long sequence data. Thus, the running time of the compared methods were compared under different parameter settings. We increased the minimum utility threshold from $\delta_1$ to $\delta_6$ on each dataset while keeping the tested data size fixed. To obtain accurate experimental results under each setting, each compared approach was ran three times, and the average running times are plotted in Figure \ref{fig:Runtime}. As shown, the runtime of USpan exceeded 10,000 seconds in Leviathan when the minimum utility threshold was lower than 1.20\%, and thus USpan only has two points in Figure \ref{fig:Runtime}(e).

As shown in each sub-figure of Figure \ref{fig:Runtime}, ProUM is intuitively the most efficient except in yoochoose-buys. Both the proposed algorithm with or without using the PUK strategy (ProUM and ProUM$^*$) to prune the unpromising candidates before constructing the utility-arrays consistently outperformed the state-of-the-art HUS-Span approach, even by up to 3 orders of magnitude. For the Kosarak10k data in Figure \ref{fig:Runtime}(d), the performance of ProUM decreased when $\delta$ increased, but it decreased slowly afterwards when $\delta$ = 1.72\%. USpan always required a longer execution time than ProUM and ProUM$^*$, from 2,890 seconds to 130 seconds. In particular, HUS-Span had the longest execution time in this dataset, and it consumed 5,000 seconds when $\delta$ was smaller than 1.71\%. In general, ProUM outperformed ProUM$^*$ in all the test datasets under different parameter settings. For example, in Figure \ref{fig:Runtime}(a), the difference of the runtime  between ProUM$^*$ and ProUM can be observed. When $\delta$ = 1.2\% on the Sign dataset, the runtime of ProUM$^*$ closed to 115 seconds while the runtime of ProUM was approximately 60 seconds.  These observations can also be intuitively seen on other datasets, such as Figure \ref{fig:Runtime}(c), Figure \ref{fig:Runtime}(d), and Figure \ref{fig:Runtime}(e). These observations indicate that the \textit{local downward closure} property of the \textit{SEU} upper bound plays an active role in pruning the search space of the projection-based ProUM algorithm. 

In addition, it is interesting to observe that USpan sometimes ran even faster than HUS-Span. In many cases, however, it is not clear whether the recently proposed HUS-Span was faster than the USpan method that was optimized for this experiment. For example, when the experiment was conducted on Sign, SynDataset-160K, and Kosarak10k, it seems that HUS-Span had a longer running time than USpan. For the SynDataset-160K shown in Figure \ref{fig:Runtime}(c), HUS-Span was the most time consuming among the four algorithms. It required 2,824 seconds when $\delta$ was set to 0.065\%, which was quite a bit longer than the others. The performance of USpan, ProUM$^*$, and ProUM declined before $\delta = 0.075\%$, and it nearly remained stable afterwards. In other datasets, such as Bible, Leviathan, and yoochoose-buys, USpan performed worse than HUS-Span as well as the two variants of the proposed ProUM model. A possible reason for this is that HUS-Span needs additional time to scan the projected sub-databases for calculating the utility information from the built utility-chains. In addition, in some datasets, the upper bound \textit{PEU} sometimes had a similar effect to that of the proposed \textit{SEU} upper bound.

The projection mechanism of utility-array makes a contribution to the improvement, which can be observed in SynDataset-160K and Kosarak10k. This is because the small size of the utility-array creates a favorable \textit{SEU} value that enhances the computation of the later processes. Specifically, based on an observation of the runtime between ProUM and ProUM$^*$, the PUK strategy plays an active role in filtering the unpromising patterns before constructing the set of utility-arrays. In summary, the enhanced ProUM algorithm that utilizes powerful pruning strategies always had the best performance compared to the baseline ProUM$^*$ as well as USpan and the state-of-the-art HUS-Span algorithm. The designed ProUM algorithm is acceptable and efficient in discovering high-utility sequential patterns on different types of datasets.

\textbf{Summary of efficiency study}. The above-stated results demonstrate the efficiency of ProUM. Under different parameter settings (when $\delta$ is large), ProUM always required less time than the existing HUSPM algorithms. In addition, the divide-and-conquer strategy was applied to project the utility-arrays during the recursive mining processes. All the experimental results demonstrate the suitability of the proposed ProUM models for dealing with both real or synthetic datasets.

\begin{figure*}[htbp]
	\centering 
	\includegraphics[trim=0 5 10 0,clip,scale=0.56]{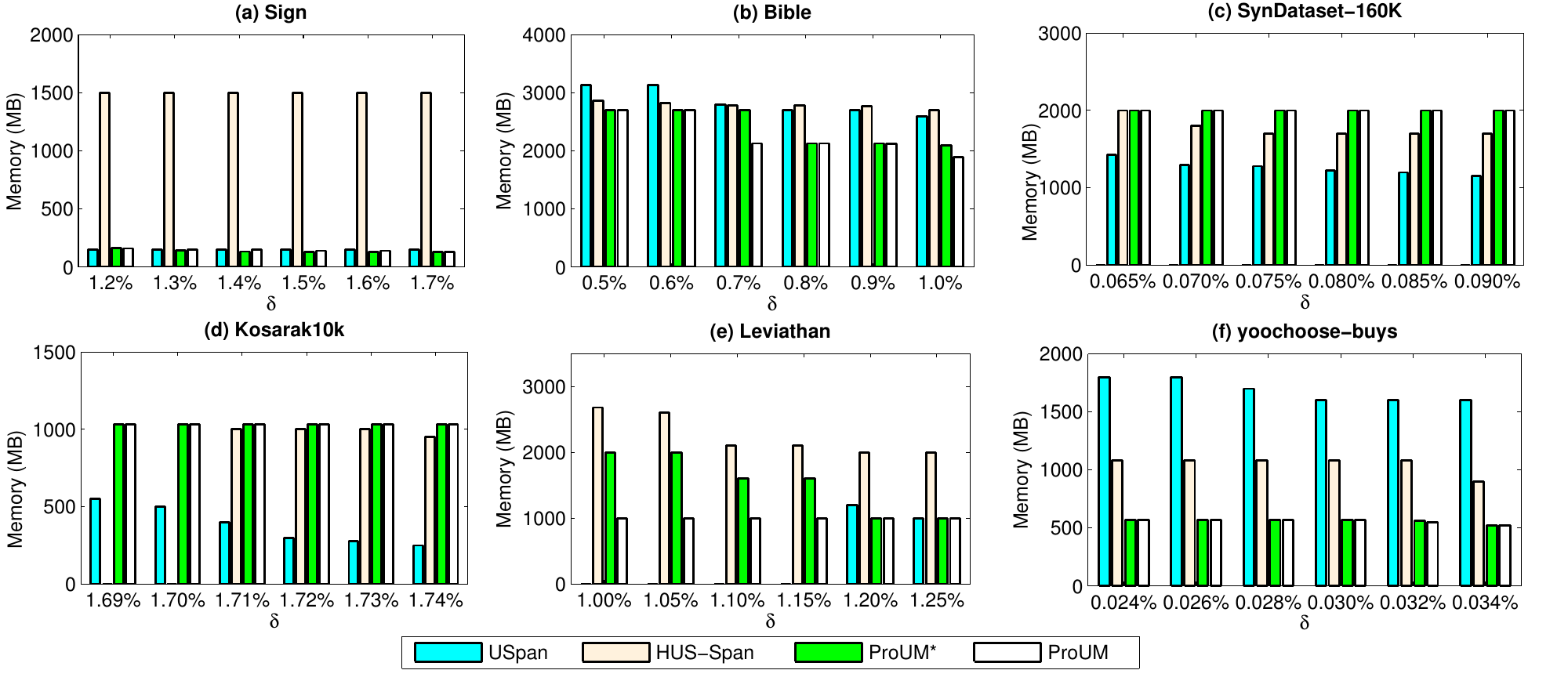}
	\captionsetup{justification=centering}
	\caption{Memory usage by varying $\delta$}
	\label{fig:Memory}	
\end{figure*}

\subsection{Candidate Analysis}

The generated patterns of the four compared algorithms were investigated to evaluate the effect of pruning strategies by conducting under the same parameter settings, as shown in Figure \ref{fig:Runtime}. The results of the different kinds of generated patterns, both generated candidates and final HUSPs, are plotted in Table \ref{table:pattern}. Note that the \#HUSPs is the number of final discovered HUSPs, and \#\textit{P}1, \#\textit{P}2, \#\textit{P}3, and \#\textit{P}4 are the numbers of the candidates generated by USpan, HUS-Span, ProUM$^*$, and ProUM, respectively.

As shown in Table \ref{table:pattern}, it can be clearly observed that, on all the tested datasets, the number of HUSPs was always quite a bit less than that of the candidate patterns (e.g., \#\textit{P}1, \#\textit{P}2, \#\textit{P}3, and \#\textit{P}4) under various minimum utility thresholds. For example, in Kosarak10k, the discovered HUSPs were changed from 23 to 21 while the related candidates were increased from 8,753,634 up to 23,917,337. These results reflect the fact that there are a huge number of candidate patterns that are generated in a HUSPM algorithm but very few of them are the final interesting desired patterns. As mentioned previously, there are several challenges in utility mining when dealing with sequence data. How to effectively prune the search space in HUSPM is more difficult due to the absence of the downward closure property in the sequence utility.

Intuitively, on all tested datasets with different parameter settings, \#\textit{P}1 and \#\textit{P}4 was nearly equal to \#\textit{P}2 while \#\textit{P}2 had the most number of candidates among all the compared candidate patterns. This is because the previously mentioned upper bound error in the USpan algorithm was replaced by the proposed \textit{SEU} upper bound in our experiments. Thus, both USpan and ProUM used the same upper bounds, \textit{SWU} and \textit{SEU}, to prune the search space. This results in nearly the same results of the number of the candidate patterns. It is interesting to observe that \#\textit{P}4 was nearly equal to \#\textit{P}2, which indicates that the \textit{SEU} used in ProUM had a similar powerful pruning ability to \textit{PEU} that was used in HUS-Span for the addressed HUSPM problem.

Specifically, the difference between \#\textit{P}3 and \#\textit{P}4 
proves the effectiveness of the PUK strategy for ProUM. That is, the proposed \textit{SEU} upper bound has a better ability to prune the search space than the loose \textit{SWU}. Although both \textit{SWU} and \textit{SEU} affect the candidate patterns for mining HUSPs, in general, the numbers of \#\textit{P}4 were always quite smaller than those of \#\textit{P}3. For example, as shown in Bible, \#\textit{P}4 changed from 262,465 to 42,205 while \#\textit{P}3 had its number decreased from 100,706 to 17,831 when varying $\delta$ from 0.5\% to 1.0\%. Therefore, ProUM$^*$ adopts the PUO strategy to filter the unpromising items in \textit{iItem}, and \textit{sItem} is not more powerful than ProUM, which utilizes PUK strategy (w.r.t. the \textit{SEU} upper bound) to remove these unpromising items. 

\textbf{Discussion}. These results of the patterns indicate that the proposed upper bound \textit{SEU} is more suitable than \textit{SWU} to prune the subtrees of LQS-tree for mining HUSPs. The results demonstrate the positive effect of pruning strategies in ProUM for discovering utility-driven sequential patterns.

\subsection{Memory Evaluation}

In this subsection, the mining efficiency is evaluated in terms of memory consumption. All parameters are set to the default values shown in Figure \ref{fig:Runtime} unless otherwise stated. Figures \ref{fig:Memory}(a) to (f) respectively show plots of the results of the peak memory usage of all the compared algorithms. Note that Java API was used to calculate the peak memory consumption of each compared algorithm during the whole mining process.

As shown, the projection utility-array-based models, both ProUM$^*$ and ProUM, performed significantly better than the baselines. Although the  HUS-Span algorithm also utilizes the projection technique, it needed to project the sub-databases before the construction of utility-chains. This consumes more execution time and memory cost than ProUM, which only projects and scans the sub-utility-arrays. For example, as shown in Figures \ref{fig:Memory} (a) and (f), the peak memory consumption for ProUM was significantly less than that of HUS-Span because ProUM consumes the reasonable memory to store the compact utility-arrays and generate promising patterns. In addition, the improved variant ProUM consumes less memory than the baseline ProUM$^*$ that adopts the PUO strategy to remove the unpromising items.

Figure \ref{fig:Memory} shows the effects of the parameter - minimum utility threshold $\delta$ on the memory performance of ProUM. As shown on all datasets, the memory usage of ProUM$^*$ and ProUM did not change much when $\delta$ increased while the memory usage of USpan and HUS-Span may change more substantially in most cases. For example, when $\delta$ = 1.20\%, ProUM consumed only 1,000 MB on Leviathan while HUS-Span consumed more than 2,000 MB. The performance gap is more obvious on yoochoose-buys, mainly because USpan and HUS-Span were ineffective on Leviathan and yoochoose-buys. The decrease with $\delta$ was quite rapid because a small $\delta$ makes all the compared HUSPM algorithms execute searching in LQS-tree more times, which makes it harder to return the mined results, especially for the existing USpan and HUS-Span algorithms. In addition, a very large $\delta$ brings no extra benefit to the discovered results of HUSPM. Hence, in practice, $\delta$ does not need to be too large.

\textbf{Summary}. The proposed ProUM model with several pruning strategies consumed less memory than HUS-Span for all the parameter settings and also less than that of the optimized USpan in most cases. USpan had the least memory consumption in Figures \ref{fig:Memory}(c) and \ref{fig:Memory}(d). Nonetheless, the best performing USpan on memory consumption for these cases had much worse execution times in Figure \ref{fig:Runtime}(d). As previously mentioned, one of the advantages of ProUM is that it is able to filter a large amount of unpromising patterns at an early stage by building the projected utility-arrays.

\subsection{Scalability Test}

\begin{figure*}[!htbp]
	\setlength{\abovecaptionskip}{0pt}
	\setlength{\belowcaptionskip}{0pt}	
	\centering
	\includegraphics[trim=30 200 50 0,clip,scale=0.58]{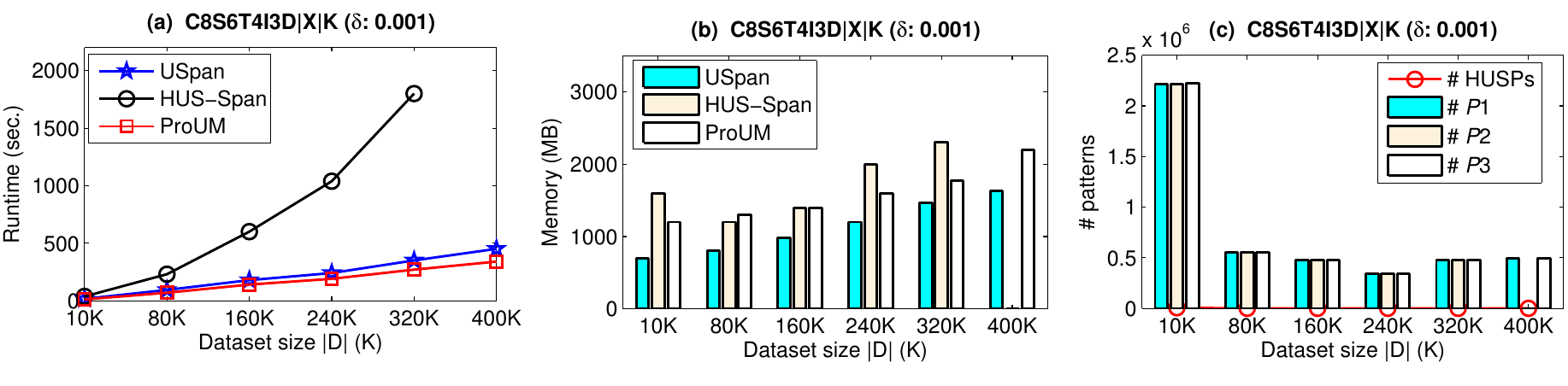}
	\caption{Scalability test.}
	\label{fig:scalability}
\end{figure*}

Scalability is important mainly because many real-world data is massive, especially large-scale sequence data. Therefore, scalability is an important acceptance criteria for a designed data mining model. In this subsection, ProUM is analytically compared with the existing methods on a large-scale dataset, and the experimental results are shown in Figures \ref{fig:scalability} (a) to (c), respectively.

Figure \ref{fig:scalability} shows the results of scalability performance in the synthetic dataset C8S6T4I3D$|$X$|$K in terms of different data sizes from 10K to 400K sequences. The running time of each algorithm increased linearly as the number of sequences grew. ProUM showed superior scalability with respect to the dataset size among the compared methods. For the test dataset that contained element-based sequences, USpan performed better than the HUS-Span algorithm that utilizes the \textit{PEU} upper bound. For example, the running time of HUS-Span exceeded 2,000 seconds when dealing with 400K sequences. According to the memory usage, ProUM required more memory than other baselines, as shown in the center sub-figure. In addition, the amount of memory used was bounded in ProUM through the pruning strategies that were employed. The candidate patterns still show that \#\textit{P}3 was similar to \#\textit{P}1 and \#\textit{P}2 (Figure \ref{fig:scalability}(c)). Note that here \#\textit{P}3 is number of generated candidates in ProUM.

\textbf{Discussion}. In summary, the scalability results confirm the intuition that the projected ProUM method using utility-array with a series of indexing positions is scalable for large-scale datasets, and it is superior to the existing algorithms.

\section{Conclusions} 
\label{sec:conclusion}

In general, utility-based sequence analytics are more useful than other support-based data mining techniques. However, utility mining on sequence data can easily suffer from several problems, not just with critical combinational explosion but also from computational complexity caused by sequencing between itemsets/elements. How to improve the mining efficiency of utility mining on sequence data is still an open problem. To this end, we developed a projection-based utility mining algorithm, called ProUM, for the fast mining of high-utility sequential patterns. A new data structure, called utility-array, was also proposed, which can be directly used to calculate the utility and remaining utility of a sequence without scanning the database. Based on the projection mechanism in applying in utility-array, the presented solutions, including two utility bounds, the corresponding pruning strategies, and the ProUM algorithm, are proposed. ProUM was compared with USpan and HUS-Span, which are the state-of-the-art algorithms for mining HUSPs in sequence data. Extensive experimental results on both synthetic and real-life datasets  demonstrated that ProUM had a better efficiency compared to the state-of-the-art baselines.

\section*{Acknowledgment}
This work was partially supported by the Shenzhen Technical Project under project No. KQJSCX 20170726103424709 and No. JCYJ 20170307151733005. Specifically, Wensheng Gan was supported by the CSC (China Scholarship Council) Program during the study at University of Illinois at Chicago, IL, USA.

\section*{References}
\bibliography{paper} 

\end{document}